\documentclass[aps,pra,twocolumn,amsmath,amssymb,bibnotes]{revtex4}

\usepackage{color,epsfig,rotating,graphicx,subfigure}

\usepackage[latin1]{inputenc}
\usepackage[english]{babel}

\usepackage{dsfont}
\usepackage{textcomp}
\usepackage{tikz}
\usepackage{mathtools}
\usepackage{color}

\renewcommand{\Im}{{\rm Im}}

\newcommand{\ri}{{\rm i}}

\newcommand{\rd}{{\rm d}}

\newcommand{\chimat}{\underline{\underline{\chi}}}

\newcommand{\blockt}{\boldsymbol{T}^{-1}}
\newcommand{\tildeblockt}{\tilde{\boldsymbol{T}}^{-1}}
\newcommand{\I}{{\rm i}}

\newcommand{\IM}{{\rm Im \,\,}}

\begin{document}
	\title{{Thermal} near-field energy density and LDOS in topological 1D SSH chains and 2D SSH lattices of plasmonic nanoparticles}
	
	\date{\today}
	
	\author{Annika Ott$^1$, Zhenghua An$^2$, Achim Kittel$^1$, and Svend-Age Biehs$^{1,*}$}
	\affiliation{1 Institut f{\"u}r Physik, Carl von Ossietzky Universit{\"a}t, D-26111 Oldenburg, Germany}
	\affiliation{2 State Key Laboratory of Surface Physics, Department of Physics and Institute for Nanoelectronic Devices and Quantum Computing, Fudan University, Shanghai 200433, China}
	
	\email{ s.age.biehs@uni-oldenburg.de} 
	
	
	\begin{abstract}
		We derive a general expression for electric and magnetic part of the near-field energy density of $N$ dipoles of temperatures $T_1, \ldots, T_N$ immersed in a background field having a different temperature $T_b$. In contrast to former expressions this inclusion of the background field allows for determining the energy density of heated or cooled isotropic dipolar objects within an arbitrary environment which is thermalized at a different temperature. Furthermore, we show how the energy density is related to the local density of states. We use this general expression to study the near-field enhanced energy density at the edges and corners of 1D Su-Schrieffer-Heeger chains and 2D Su-Schrieffer-Heeger lattices of plasmonic InSb nanoparticles when the phase transition from a topological trivial to a topological non-trivial state is made. We discuss the robustness of these modes when adding defects and the possibility to measure the topological edge and corner modes. 
	\end{abstract}
	
	\maketitle

	
	\section{Introduction}
	
	Today, it is a well-established fact that the thermal energy density is enhanced in the vicinity of materials which support
	surface modes. This effect was already found in early works of Eckhardt~\cite{Eckhardt82} and then later has been studied in great detail
	within the framework of fluctuational electrodynamics as reviewed by Dorofeyev and Vinogradov~\cite{Dorofeyev}, for instance. The interest
	in the near-field energy density is caused by its connection to the local density of states (LDOS)~\cite{Eckhardt82,Dorofeyev,Joulain} {which determines the emission rates of atoms and molecules and, last but not least, it is of high relavance for near-field thermal measurements utilizing tip-based probes~\cite{Dorofeyev97,Dorofeyev98,JPendry99,MuletEtAl01,Joulain2,Jarzembski,Herz2018,Herz2021}. These measurements are for example the}
	near-field scanning thermal microscope (NSThM)~\cite{Kittel2005,Kittel2008,WorbesEtAl2013}, the thermal radiative scanning tunneling 
	microscope (TRSTM)~\cite{DeWilde,Babuty}, the  thermal infrared near-field spectroscope (TINS)~\cite{Huth2011,Jones, OCallahan}, and the scanning 
	noise microscope (SNoiM)~\cite{Lin,WengEtAl2018, Komiyama}. 
	
	In this work, we generalize the expression for the near-field energy density in a many-body system consisting of $N$ dipolar objects~\cite{TervoED} in order to study it in the near-field of topological many-body systems. In the last ten years, the investigation of near-field thermal radiation in such many-body systems based on the strict formalism of fluctuational electrodynamics~\cite{PBAmanybody,Messina2011,KruegerEtAl2012,nteilchen,DongEtAl2017,LatellaEtAl2017} has shed some light on specific many-body effects like  ballistic, diffusive, sub- and superdiffusive transport regimes~\cite{PBAEtAl2013,LatellaEtAl2018,Kathmann,Tervo,Tervo2}, heat flux switching~\cite{Incardone,Nikbakht}, persistent heat currents and heat fluxes~\cite{zhufan,zhufan2,Silveirinha,meinpaper}, persistent spins and angular momenta~\cite{meinpaper,Zubin2019}, giant magneto-resistance~\cite{Latella2017,Cuevas,He2020}, the Hall effect for thermal radiation for magneto-optical materials and topological Weyl semi-metals~\cite{hall,OttEtAl2019,OttEtAl2020}, and the dynamical control via magneto-optical surface waves~\cite{paper_diode} as reviewed recently in Refs.~\cite{Carlos,RMP}. Our specific interest is to study the energy density in the vicinity of topological 1D Su-Schrieffer-Heeger (SSH) chains~\cite{Weick1,Weick2,OESSH,ACSphotonSSH,JAPSSH} and 2D SSH lattices~\cite{sshgitter,2dzakphase,quellec} of plasmonic InSb nanoparticles (NPs) when they are undergoing a topological phase transition. Our findings are not only of fundamental interest but also hint to the possibility to measure this phase transition and in particular the topological edge and corner modes with near-field thermal probes.
	
	Our work is organized as follows: We start in Sec.~II with the derivation of a generalized expression of the near-field energy density for N dipolar objects within a given background of different temperature than the NPs. In Sect.~III we introduce the SSH chain and discuss the band structure and the appearance of topological edge modes. To this end we introduce the Zak phase. In Sec.~IV we introduce the 2D SSH model and again its band structure and appearance of topological edge and corner modes. In this case we calculate the Zak phase by means of a Wilson loop.  Finally, in Sec.~V we discuss the numerical results for the energy density in the near-field of a 1D SSH chain and a 2D SSH lattice and the impact of the topological edge and corner modes as well as the robustness of the results with respect to defects. 
	
	
	\section{energy density and LDOS}
	
	We start with the derivation of the energy density in the vicinity of a configuration of plasmonic NPs using the dipole approximation. This quantity has for example already been considered by Tervo {\itshape et al.} to study the near-field energy density of 0D, 1D, 2D, and 3D NP systems~\cite{TervoED}. {There, it was only considered the near-¿eld energy density generated by the thermal sources within the NPs. Here we also include the possibility to assign a temperature to the thermal photons in the background of the NPs. This can be very generally valid as long as it is given in terms of the Green function and as long as the NPs are themselves within the vacuum part of the environment.} Hence our approach can handle situations where a set of NPs is close to a substrate, in a cavitiy, etc. For convenience we focus only on an infinitely large vacuum background later. Furthermore, we include the magnetic part of the energy density which is negligible in the system considered in Ref.~\cite{TervoED} and therefore has not been considered.  
	
	Let us start with the electric and magnetic fields at position $\mathbf{r}_0$ generated by the fluctuational dipole moments of $N$ NPs at positions $\mathbf{r}_i$ and a background field (superindex $b$) given by
	\begin{align}
	\mathbf{E}(\omega,\mathbf{r}_0) &= \mu_0\omega^2\sum_{i=1}^{N} \mathds{G}^{\rm EE}_{0i}\mathbf{p}_i+\mathbf{E}_0^b, \label{eqg}\\
	\mathbf{H}(\omega,\mathbf{r}_0) &= \mu_0\omega^2\sum_{i=1}^{N} \mathds{G}^{\rm HE}_{0i}\mathbf{p}_i+\mathbf{H}_0^b,
	\end{align}
	where $\mathbf{p}_i = \mathbf{p}^{\rm ind}_i + \mathbf{p}^{\rm fl}_i$ consists of a purely fluctuational dipole moment $ \mathbf{p}^{\rm fl}_i$ due to the thermal agitation of charges inside the NPs and an induced dipole moment $\mathbf{p}^{\rm ind}_i$. Note that for the Green functions we use a short-hand notation $ \mathds{G}^{\rm EE}_{0i} = \mathds{G}^{\rm EE}(\mathbf{r}_0,\mathbf{r}_i)$ and $\mathds{G}^{\rm HE}_{0i} = \mathds{G}^{\rm HE} (\mathbf{r}_0,\mathbf{r}_i)$. These two Green functions are related to each other by $\mathds{G}^{\rm HE}_{0i} = (\ri \omega \mu_0)^{-1} \nabla_{\mathbf{r}_0} \times \mathds{G}^{\rm EE}_{0i}$ as dictated by the Faraday law $\nabla \times \mathbf{E} = \ri \omega \mu_0  \mathbf{H}$. Here $\mu_0$ is the permeability of vacuum. 
	
	With these fields we can calculate the mean energy density given by
	\begin{eqnarray}
	\langle u \rangle =\frac{\epsilon_0}{2}\langle |\mathbf{E}(\mathbf{r},t)|^2\rangle+\frac{\mu_0}{2}\langle |\mathbf{H}(\mathbf{r},t)|^2 \rangle,
	\label{u}
	\end{eqnarray}
	where $\epsilon_0$ is the permittivity of vacuum. To this end, we need the correlation functions of the fluctuational dipole moments and the fluctuational background fields. Assuming that the NPs and the background are in local thermal equilibrium these correlation functions follow from the fluctuation-dissipation theorem and are given by~\cite{nteilchen} 
	\begin{equation}
	\langle \mathbf{p}^{\rm fl}_k \otimes \mathbf{p'}^{\rm fl^*}_n \rangle = 2\pi\delta(\omega-\omega')\delta_{kn}\frac{2\epsilon_0}{\omega}\Theta_k\chimat
        \label{Eq:FDTp}
	\end{equation}
	and~\cite{Agarwal,Eckhardt}
	\begin{equation}
	\langle \mathbf{A}^{\rm b}_k \otimes \mathbf{B'}^{\rm b^*}_n \rangle = 2\pi\delta(\omega-\omega') 2 \omega \mu_0\Theta_b \frac{\mathds{G}_{kn}^{AB}-\mathds{G}_{nk}^{BA^\dagger}}{2\I}
          \label{Eq:FDTF}
	\end{equation}
	with $A,B\in\{E,H\}$, $\mathbf{A},\mathbf{B}\in\{\mathbf{E},\mathbf{H}\}$ and 
	\begin{equation}
        	\Theta_k= \frac{\hbar\omega}{\exp(\hbar\omega)/(k_B T_k)-1}. 
	\end{equation}
	Here we have introduced the Boltzmann constant $k_B$ and the susceptibility of the NPs which is related to the polarizability $\alpha$ of the isotropic NPs and is in our case [see also Eq.~(\ref{Eq:Ci})]~\cite{RMP}
	\begin{equation}
		\chimat = \biggl(\Im(\alpha) - \frac{k_0^3}{6 \pi} |\alpha|^2\biggr) \mathds{1} = \chi \mathds{1}. 
	\end{equation}
	Here  $k_0 = \omega/c$ is the vacuum wave number, $c$ is the light velocity in vacuum.
	Moreover, we assume that the fluctuational dipole moments $\mathbf{p}_i^{\rm fl}$ of different NPs and the background fields are statistically independent. Finally, we have neglected the vacuum fluctuations in the above expressions focusing only on the thermal contributions.
	
	Note that the different Green functions are for example defined in Ref.~\cite{Eckhardt}. Assuming that the environment is reciprocal we have the symmetry relations ${\mathds{G}^{\rm EE}_{ij}}^{t}=\mathds{G}^{\rm EE}_{ji}$, ${\mathds{G}^{\rm HH}_{ij}}^t=\mathds{G}^{\rm HH}_{ji}$,  and ${\mathds{G}^{\rm EH}_{ij}}^t = -\mathds{G}^{\rm HE}_{ji}$. So that we can express the mean energy density solely by $\mathds{G}^{\rm E} \equiv \mathds{G}^{\rm EE}$, $\mathds{G}^{\rm H} \equiv \mathds{G}^{\rm HE}$, and $\mathds{G}^{\rm HH}$. Moreover, we use
	\begin{eqnarray}
		\chi\blockt(\blockt)^\dagger&=&-k_0^2|\alpha|^2\blockt\IM\mathds{G}^{\rm E}(\blockt)^\dagger \notag \\
	& &+\frac{\blockt\alpha-(\blockt)^\dagger \alpha^*}{2\ri}
	\label{eqrelation}
	\end{eqnarray}	
	which is explicitly derived in appendix \ref{relation} with	
	\begin{equation}
	  \boldsymbol{T}_{ij} = \delta_{ij}\mathds{1}-(1-\delta_{ij})k_0^2\alpha\mathds{G}_{ij}^{\rm E}.
\end{equation}
 In Eq. (\ref{eqrelation}) all matrices are block matrices, so that for a general block matrix $(\mathds{F})_{ij}=\mathds{F}_{ij}$ gives the $i,j$-th component and $(\mathds{F}^\dagger)_{ij}=(\mathds{F}_{ji})^\dagger$. 
	Using the above introduced expressions and $\mathds{F}_{ij}^\dagger\coloneqq(\mathds{F}_{ij})^\dagger$ we obtain for the mean energy density 
	\begin{eqnarray}
	\langle u \rangle =\int_{0}^{\infty}\frac{{\rm d}\omega}{2\pi}\Bigg[u_1+u_2+u_3\Bigg]
	\label{ub}
	\end{eqnarray}
	with
	\begin{equation}
	\begin{split}
	u_1&=\sum_{i,k,m=1}^{N}\Big(\Theta_k-\Theta_b\Big)\frac{k_0^3}{c} 2 \Im(\alpha)  \\
	&\quad \times {\rm ReTr} \Bigg\{\mathds{G}_{0i}^{E}\tildeblockt_{ik}(\tildeblockt)_{mk}^\dagger(\mathds{G}^{E}_{0m})^\dagger \\
	&\quad + \frac{\mu_0}{\epsilon_0} \mathds{G}_{0i}^{H}\tildeblockt_{ik}(\tildeblockt)_{mk}^\dagger(\mathds{G}_{0m}^{H})^\dagger\Bigg\} 
	\end{split}
	\end{equation}
	and
	\begin{equation}
	\begin{split}
	u_2&=\sum_{i,j=1}^{N}2\frac{k_0^3}{c}\Bigg\{\Big(\Theta_j-\Theta_b\Big)\Im(\alpha) \\
	&\quad\times 2{\rm ReTr}\Big\{\mathds{G}_{0i}^{E}\tildeblockt_{ij}(\mathds{G}^{E}_{0j})^\dagger 
	+ \frac{\mu_0}{\epsilon_0}\mathds{G}_{0i}^{H}\tildeblockt_{ij}(\mathds{G}_{0j}^{H})^\dagger\Big\} \\
	&\quad+\Theta_b{\rm ImTr}\Big\{\alpha\mathds{G}_{0i}^{E}\tildeblockt_{ij}\mathds{G}_{0j}^{E}+\alpha\frac{\mu_0}{\epsilon_0}\mathds{G}_{0i}^{H}\tildeblockt_{ij}\mathds{G}_{0j}^{H}\Big\} \Bigg\}
	\end{split}
	\end{equation}
	and
	\begin{equation}
	\begin{split}
	u_3 &=\sum_{j=1}^{N} \frac{k_0^3}{c} 2\Bigg\{\Big(\Theta_j-\Theta_b\Big)\Im(\alpha) \\
	&\quad\times {\rm ReTr}\Big\{\mathds{G}_{0j}^{E}(\mathds{G}^{E}_{0j})^\dagger+ \frac{\mu_0}{\epsilon_0} \mathds{G}_{0j}^{H}(\mathds{G}_{0j}^{H})^\dagger\Big\}  \\
	&\quad+\Theta_b{\rm ImTr}\Big\{\alpha\mathds{G}_{0j}^{E}\mathds{G}_{0j}^{E} + \alpha\frac{\mu_0}{\epsilon_0}\mathds{G}_{0j}^{H}\mathds{G}_{0j}^{H}\Big\}\Bigg\} \\
	&\quad + 2 \Theta_b \frac{k_0}{c}\biggl({\rm ImTr}\mathds{G}_{00}^{E} + {\rm ImTr}\mathds{G}_{00}^{ \rm HH} \frac{\mu_0}{\epsilon_0} \biggr).
	\end{split}
	\label{ubt}
	\end{equation}
	This expression is one of the main results of this work. The different terms $u_1$, $u_2$, $u_3$ are chosen such that they contain the matrix $\tildeblockt_{ij}$ two, one, or zero times where 
	\begin{equation}
			\tilde{\boldsymbol{T}}_{ij}^{-1} = \boldsymbol{T}_{ij}^{-1} - \mathds{1}.   
	\end{equation}

	This derived expression for the spectral energy density
	\begin{equation}
	u_\omega = u_1+u_2+u_3
	\end{equation}
	applies for any choice of temperatures $T_1, \ldots, T_N, T_b$. In particular it becomes quite simple in the case of global thermal
	equilibrium where $T_1 = T_2 = \ldots = T_N = T_b$. We obtain
{	\begin{equation}
	\begin{split}
	u_\omega^{\rm eq} &= 2 \Theta_b \frac{k_0}{c} \biggl({\rm ImTr}\mathds{G}_{00}^{E} + {\rm ImTr}\mathds{G}_{00}^{ \rm HH} \frac{\mu_0}{\epsilon_0} \biggr)\\
	&\quad + \sum_{j = 1}^N 2 \Theta_b \frac{k_0^ 3}{c}{\rm ImTr}\Big\{\alpha\mathds{G}_{0j}^{E}\mathds{G}_{0j}^{E} + \alpha\frac{\mu_0}{\epsilon_0}\mathds{G}_{0j}^{H}\mathds{G}_{0j}^{H}\Big\} \\
	&\quad + \sum_{i,j = 1}^N 2 \Theta_b  \frac{k_0^ 3}{c} {\rm ImTr}\Big\{\alpha\mathds{G}_{0i}^{E}\tildeblockt_{ij}\mathds{G}_{0j}^{E}+\alpha\frac{\mu_0}{\epsilon_0}\mathds{G}_{0i}^{H}\tildeblockt_{ij}\mathds{G}_{0j}^{H}\Big\} .
	\end{split}
	\end{equation}
	From this expression the LDOS $D (\omega,\mathbf{r}_0)$ follows directly since $D (\omega,\mathbf{r}_0) = u_\omega^{\rm eq} / (2 \pi \Theta_b)$. The energy density and LDOS consist obviously of three parts: the direct part of the environment (first term), the direct part of the NPs (second term) which is proportional to $\alpha$, and the scattering term (third term) which is proportional to $\alpha^2$ . }
	
        In the following we focus on the energy density in the vicinity of a 1D SSH chain and a 2D SSH lattice in a pure vacuum background so that the Green functions are those of the situation in bare vacuum, but other environments can be taken into account by using the corresponding expressions of the Green functions. The vacuum electric Green function $\mathds{G}_{ij}^{\rm E} \coloneqq\mathds{G}^{\rm E}(\mathbf{r}_{i},\mathbf{r}_{j})$ is given by~\cite{Novotny}
	\begin{equation}
	  \mathds{G}^{\rm E}_{ij}=(a\mathds{1}+b\mathbf{e}\otimes\mathbf{e})\frac{e^{\I k_0\tilde{d}_{ij}}}{4\pi\tilde{d}_{ij}} 
	\end{equation}
	with 
	\begin{align}
          \tilde{d}_{ij}&=|\mathbf{r}_{i}-\mathbf{r}_{j}|,\\
  	  a &= 1+\frac{ik_0\tilde{d}_{ij}-1}{k_0^2\tilde{d}_{ij}^2}, \\
	  b &=\frac{3-3ik_0\tilde{d}_{ij}-k_0^2\tilde{d}_{ij}^2}{k_0^2\tilde{d}_{ij}^2}, \\
	\mathbf{e}&=(\mathbf{r}_{i}-\mathbf{r}_{j})/\tilde{d}_{ij}
	\end{align}
         and the magnetic Green function is given by ~\cite{Novotny}
	\begin{eqnarray}
		\mathds{G}_{ij}^{H}=\frac{e^{\I k_0\tilde{d}_{ij}}}{4\pi\tilde{d}^2_{ij}}\frac{l_{ij}}{\mu_0c}\begin{pmatrix}
	0 & z_j-z_i &y_i-y_j\\z_i-z_j&0&x_j-x_i\\y_j-y_i&x_i-x_j&0
	\end{pmatrix}
	\end{eqnarray}
	with 
	\begin{eqnarray}
		l_{ij} = 1+\frac{\I}{k_0\tilde{d}_{ij}}.
	\end{eqnarray} 
        Furthermore, we have in the vacuum region $\mathds{G}^{\rm HH}_{ij} = \mathds{G}^{\rm EE}_{ij} \epsilon_0/\mu_0$. 
	Then from the general expressions the energy density for the longitudinal or 
	transversal modes in the SSH chain can be obtained by restricting $\mathds{G}^E_{ij}$ only 
	to the $xx$ or $zz$ component. In the same manner for the SSH lattice the energy density of the 
	ip and op polarized modes can be obtained by restricting $\mathds{G}^E_{ij}$ to
	the x-y subspace as in Eq.~(\ref{Eq:Greenxy}) or to the zz component, only.

	\section{SSH chain}

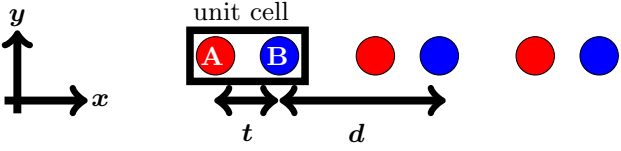
\begin{figure}[h!]
		\hspace{-0.5 cm}
		\parbox{0.5\textwidth}{
			\centering	
			\begin{tikzpicture}[transform shape,scale=0.85]
			\filldraw [above=17 mm, right=18mm, fill=red ,draw=black] (4,0) circle (3 mm);
			\filldraw [above=17 mm, right=28mm, fill=blue ,draw=black] (4,0) circle (3 mm);
			
			\draw[<->, line width=3pt, draw=black ] (5.78,1.) -- (6.78, 1.);
			\node at (6.3,0.5,0.0) {\textcolor{black}{\scalebox{1.3}{$\boldsymbol{t}$}}};		
			\draw[<->, line width=3pt, draw=black ] (6.8,1.) -- (9.4, 1.);			
			\node at (8,0.5,0.0) {\textcolor{black}{\scalebox{1.3}{$\boldsymbol{d}$}}};	
			\filldraw [above=17 mm, right=43mm, fill=red ,draw=black] (4,0) circle (3 mm);
			\filldraw [above=17 mm, right=53mm, fill=blue ,draw=black] (4,0) circle (3 mm);
			
			\draw [draw=black, line width=1mm] (5.4,1.3)  rectangle  ++(1.8,0.8);
			\node at (6.2,2.4,0.0) {\textcolor{black}{\scalebox{1.3}{unit cell}}};	
			\filldraw [above=17 mm, right=68mm, fill=red ,draw=black] (4,0) circle (3 mm);
			\filldraw [above=17 mm, right=78mm, fill=blue ,draw=black] (4,0) circle (3 mm);
			
			%
			\draw[->, line width=3pt, draw=black ] (2.5,1.) -- (3.8, 1.);
			\draw[->, line width=3pt, draw=black ] (2.7,0.8) -- (2.7, 2.1);
			\node at (2.7,2.3,0.0) {\textcolor{black}{\scalebox{1.3}{$\boldsymbol{y}$}}};	
			\node at (4.,1.,0.0) {\textcolor{black}{\scalebox{1.3}{$\boldsymbol{x}$}}};
			
			\node[above=17 mm, right=14.5mm] at (4.0,0.0,0.0) {\textcolor{white}{\scalebox{1.3}{\textbf A}}};
			\node[above=17 mm, right=24.8mm] at (4.0,0.0,0.0) {\textcolor{white}{\scalebox{1.3}{\textbf B}}};
			\end{tikzpicture}	
		}
		\caption{Cutout of an infinite SSH chain with $t=\frac{\beta}{2}d$ and lattice constant $d$. Adjacent particles belonging to the same sublattice A or B are separated by distance $d$.} 
		\label{schema_sshkette}
	\end{figure}

	
Now, we consider a chain of NPs as sketched in Fig.~\ref{schema_sshkette}. It consists of identical spherical isotropic NPs with a lattice constant $d$. The NPs A and B each are on their own sublattice. They form a unit cell with a spacing of $t=\beta d/2$ between the A and B NPs. We consider spherical InSb NPs with radius $R$ (we use $R=100$ nm) having a polarizability in the quasi-static limit ($R k_0 \ll 1$) given by
	\begin{equation}
	\alpha = 4\pi R^3\frac{\epsilon-1}{\epsilon + 2}.
	\label{Eq:alpha}
	\end{equation}
	By using this expression we neglect radiative damping and dynamical depolarization corrections~\cite{Meier1983,JensenEtAl1999}. This can be safely done if $k_0 R \ll 1$~\cite{Meier1983,JensenEtAl1999}. In our case $R = 100\,{\rm nm}$ and $\omega \approx 2 \times 10^{14}\,{\rm rad/s}$ which yields $k_0 R = 0.07$. We have checked numerically that the relative error due to the omission of the dynamical depolarization stays below 10\%.

	Furthermore, we are using only the dominant metallic part of the optical response modelled by the Drude permittivity~\cite{exp}
	\begin{equation}
	\epsilon = \epsilon_\infty \left(1 - \frac{\omega_{\rm p}^2}{\omega(\omega+{\rm i}\Gamma)} \right),
	\label{eps1}
	\end{equation}
	with the effective mass $m^* = 7.29\times10^{-32}$ kg, the density of the free charge carriers $n = 1.36\times10^{19}$ cm$^{-3}$, the high-frequency dielectric constant $\epsilon_\infty = 15.68$ and the damping constant $\Gamma = 1\times10^{12}\,{\rm s}^{-1}$. With these parameters the resonance frequency of the localized plasmonic modes in the InSb NP is $\omega_{\rm LP} = \omega_p \sqrt{\epsilon_\infty/(\epsilon_\infty + 2)} = 1.752\times10^{14}\,{\rm rad}{\rm s}^{-1}$, i.e.\ it clearly lies in the infrared regime around $\lambda_{\rm th} \approx 10\,\mu{\rm m}$ which is relevant for thermal radiation around room temperature. {Additionally}, it needs to be kept in mind that the dipole model for describing the NPs is valid as long as the radii $R$ of the NPs are much smaller than $\lambda_{\rm th}$ and the relative distance is larger than $3R$~\cite{Naraynaswamy2008,PBA2008,Otey,Becerril}

	The induced dipole moment $\mathbf{p}_{Ai}$ for an NP A in the i-th unit cell at position $\mathbf{r}_{Ai}$ by the dipole moments of all other NPs is given by  
	\begin{equation}
	\begin{split}
	\mathbf{p}_{Ai} &= k_0^2 \sum_{j \neq i} \alpha \mathds{G}^{\rm E}(\mathbf{r}_{Ai},\mathbf{r}_{Aj}) \mathbf{p}_{Aj} \\
	&\quad + k_0^2 \sum_{j} \alpha \mathds{G}^{\rm E}(\mathbf{r}_{Ai},\mathbf{r}_{Bj}) \mathbf{p}_{Bj},
	\end{split}
	\label{Eq:InducedDipole}
	\end{equation}
	where $i$ and $j$ run over all NPs of the sublattice. The equation which holds for the B NPs can be obtained by simply exchanging A and B
	
	For the infinite chain the eigenvalue equation with the usage of Bloch's theorem is given by \cite{meinsshpaper}
	\begin{equation}
	\mathds{M}^{\nu} \begin{pmatrix} p^{\nu}_A \\ p^\nu_B \end{pmatrix} = \frac{1}{\alpha}\begin{pmatrix} p^{\nu}_A \\ p^\nu_B \end{pmatrix}
		\label{Eq:SSH1D}
	\end{equation}
	with
	\begin{eqnarray}
	{M}^\nu_{\gamma\delta} =k_0^2\sum_{j, \gamma_i \neq \delta_j}{G}^{{\rm E},\nu}_{\gamma_{i}\delta_{j}}e^{\I \mathbf{k} \cdot (\mathbf{r}_{\delta_{i}}-\mathbf{r}_{\delta_{j}})}
	\label{mkomp}
	\end{eqnarray}
		where the two polarization $\nu = \perp, \parallel$ are implemented using the $xx,(zz)$ component in the Green function 
		$G^{{\rm E},\parallel}_{\gamma_{i}\delta_{j}}$ ($G^{{\rm E},\perp}_{\gamma_{i}\delta_{j}}$). Accordingly $p^{\nu}_{A (B)}$ is the dipole moment of the $\nu$ polarized $A (B)$ particle in a unit cell.
		More details about the eigenvalue equation can be found in Refs. ~\cite{OESSH,ACSphotonSSH,JAPSSH}.
	
	\begin{figure}[h!]
		\includegraphics[width=0.4\textwidth]{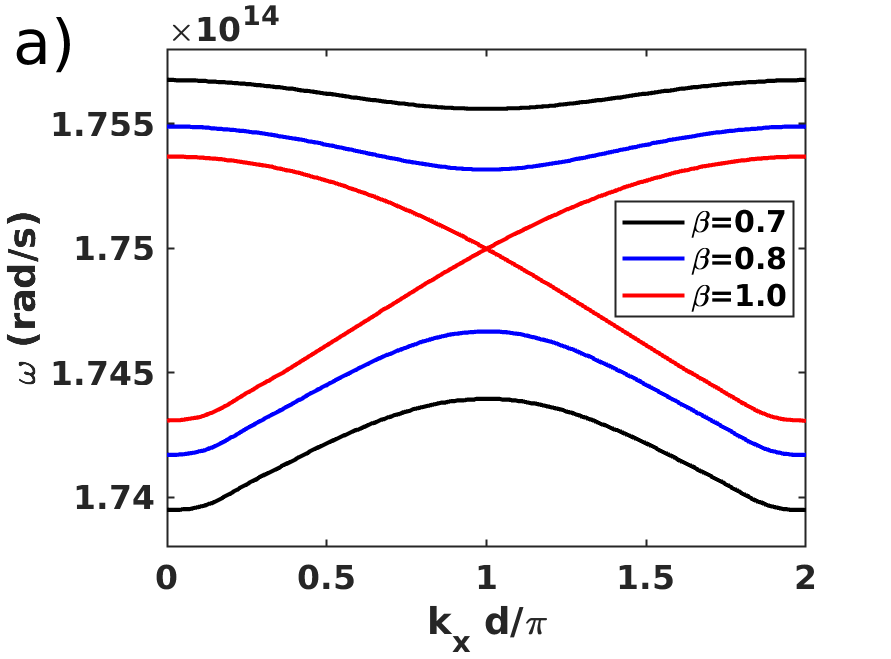}
		\includegraphics[width=0.4\textwidth]{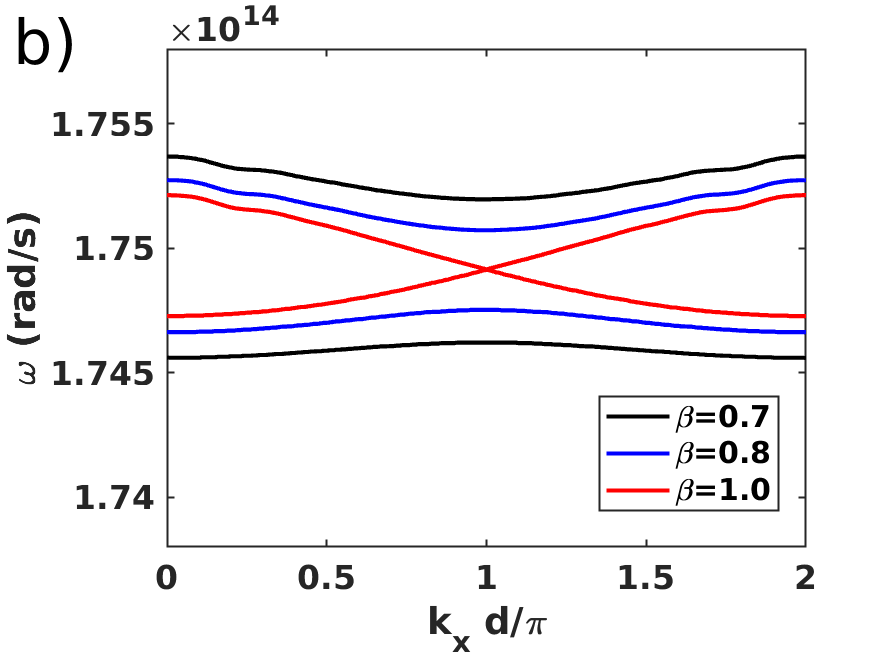}
		\caption{Bandstructure of an infinite chain with (a) longitudinal and (b) transversal polarized InSb particles with radius $R = 100\,{\rm nm}$ and lattice constant $d = 1\,\mu{\rm m}$. }
		\label{fig:Banddiagramm}
	\end{figure}

\begin{figure}[h!]
		\includegraphics[width=0.4\textwidth]{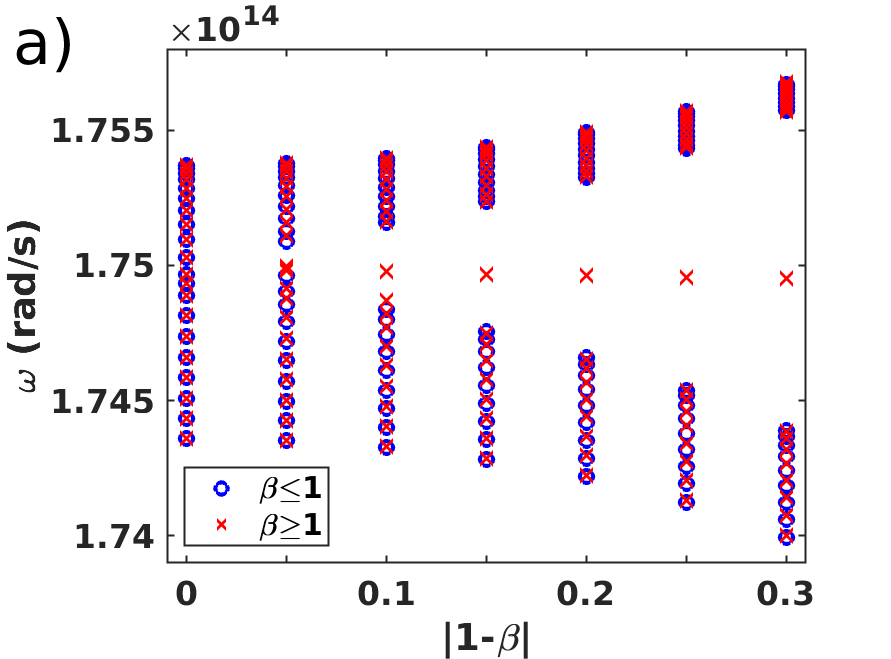}
		\includegraphics[width=0.4\textwidth]{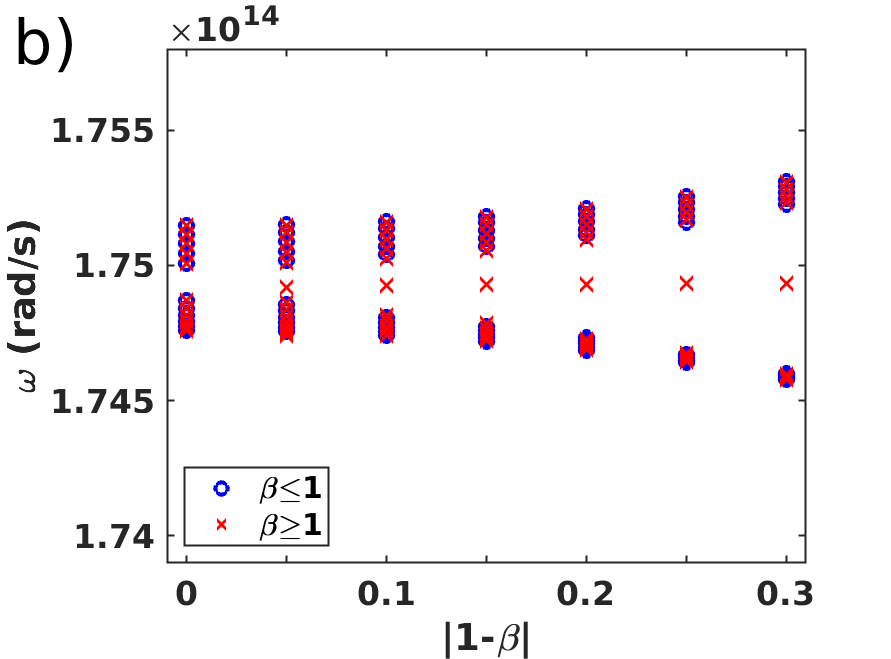}
		\caption{Real parts of the complex eigenfrequencies of the chain modes determined by Eq.~(\ref{Eq:InducedDipole}) for a) longitudinal and b) transversal polarization as a function of $\beta$. The finite chain consists of 20 InSb NPs with radius $R = 100\,{\rm nm}$ and lattice constant $d = 1\,\mu{\rm m}$. For $\beta < 1$ (blue crosses) the Zak phase is $\gamma_{\perp/\parallel} = 0$ and for $\beta > 1$ (red crosses) the Zak phase is $\gamma_{\perp/\parallel} = \pi$.}
		\label{modes}
	\end{figure}

	In Fig. ~\ref{fig:Banddiagramm} the typical two bands formed by the real part of the eigenmode frequencies of the infinite SSH chain~\cite{ford} can be seen. Furthermore, following from symmetry the bands for $\beta = 1 + x$ and $\beta = 1 - x$ with a fixed value $x$ are the same. For $\beta=1$ the band gap is closed \cite{meinsshpaper}. In addition, when taking the quasi-static regime $k_0 d \ll 1$ and $k_0 t \ll1$ into account for $\beta>1$ the Zak phase $\gamma_\nu$ is $\pi$ which is topological non trivial and for $\beta<1$ it is 0, which is topological trivial. Following from the Zak phase in the topological non trivial case for even $N$ there are two topologically protected edge modes in the band gap confined to the edges of the chain~\cite{meinsshpaper,OESSH,ACSphotonSSH,JAPSSH} as can be seen in Fig.~\ref{modes}. Of course, in the topological trivial case no edge modes can appear. Detailed calculations and discussions of the Zak phase and edge modes can be found in \cite{meinsshpaper} and are not repeated here. Note that there are cases where the Zak phase indicates a topological non-trivial state but no edge modes exist~\cite{PocockEtAl2019}. This breakdown of the bulk-edge correspondence can happen for large enough lattices constants {in the 1D and 2D case}. For the plasmonic resonance in InSb and a lattice constant of $d = 1\,\mu{\rm m}$ we find no breakdown meaning that the Zak phase clearly indicates the existence of the topological edge modes. {As discussed in Appendix~\ref{bbec} we would have a breakdown for $d \geq 3\,\mu{\rm m}$ for the transversal modes in agreement with the results in Ref.~\cite{PocockEtAl2019}.}

	Apart from the real parts of the eigenmode frequencies we have also studied the imaginary parts. We find that for the 1D SSH and the 2D SSH lattice in the next section that the imaginary parts are concentrated around the value $- \Gamma / 2$ with a small variation on the ten percent level. Therefore, the damping constant of the intensity of an excited mode is $- 2 \Im(\omega) = \Gamma$ and consequently the lifetime is approximately $1/\Gamma$ for all modes. As expected, the lifetime is mainly determined by the losses.
	
	\section{2D SSH lattice}
	
	\begin{figure}[h!]
		\hspace{-2cm}
		\parbox{0.4\textwidth}{
			\centering	
			\begin{tikzpicture}[transform shape,scale=0.9]
			\filldraw [above=17 mm, right=18mm, fill=red ,draw=black] (4,0) circle (3 mm);
			\filldraw [above=17 mm, right=28mm, fill=blue ,draw=black] (4,0) circle (3 mm);
			\filldraw [above=27 mm, right=18mm, fill=green ,draw=black] (4,0) circle (3 mm);
			\filldraw [above=27 mm, right=28mm, fill=orange ,draw=black] (4,0) circle (3 mm);
			
			\filldraw [above=42 mm, right=18mm, fill=red ,draw=black] (4,0) circle (3 mm);
			\filldraw [above=42 mm, right=28mm, fill=blue ,draw=black] (4,0) circle (3 mm);
			\filldraw [above=52 mm, right=18mm, fill=green ,draw=black] (4,0) circle (3 mm);
			\filldraw [above=52 mm, right=28mm, fill=orange ,draw=black] (4,0) circle (3 mm);
			
			\draw[<->, line width=3pt, draw=black ] (5.2,1.6) -- (5.2, 2.7);
			\node at (4.7,2.2,0.0) {\textcolor{black}{\scalebox{1.3}{$\boldsymbol{t}$}}};		
			\draw[<->, line width=3pt, draw=black ] (5.2,2.7) -- (5.2, 5.2);			
			\node at (4.8,3.85,0.0) {\textcolor{black}{\scalebox{1.3}{$\boldsymbol{d}$}}};	
			
			\filldraw [above=67 mm, right=18mm, fill=red ,draw=black] (4,0) circle (3 mm);
			\filldraw [above=67 mm, right=28mm, fill=blue ,draw=black] (4,0) circle (3 mm);
			\filldraw [above=77 mm, right=18mm, fill=green ,draw=black] (4,0) circle (3 mm);
			\filldraw [above=77 mm, right=28mm, fill=orange ,draw=black] (4,0) circle (3 mm);
			
			\filldraw [above=17 mm, right=43mm, fill=red ,draw=black] (4,0) circle (3 mm);
			\filldraw [above=17 mm, right=53mm, fill=blue ,draw=black] (4,0) circle (3 mm);
			\filldraw [above=27 mm, right=43mm, fill=green ,draw=black] (4,0) circle (3 mm);
			\filldraw [above=27 mm, right=53mm, fill=orange ,draw=black] (4,0) circle (3 mm);
			
			\draw [draw=black, line width=1mm] (7.78,1.18)  rectangle  ++(2.,2.);
			\node at (8.8,0.8,0.0) {\textcolor{black}{\scalebox{1.3}{unit cell}}};	
			
			\filldraw [above=17 mm, right=68mm, fill=red ,draw=black] (4,0) circle (3 mm);
			\filldraw [above=17 mm, right=78mm, fill=blue ,draw=black] (4,0) circle (3 mm);
			\filldraw [above=27 mm, right=68mm, fill=green ,draw=black] (4,0) circle (3 mm);
			\filldraw [above=27 mm, right=78mm, fill=orange ,draw=black] (4,0) circle (3 mm);
			
			\filldraw [above=42 mm, right=43mm, fill=red ,draw=black] (4,0) circle (3 mm);
			\filldraw [above=42 mm, right=53mm, fill=blue ,draw=black] (4,0) circle (3 mm);
			\filldraw [above=52 mm, right=43mm, fill=green ,draw=black] (4,0) circle (3 mm);
			\filldraw [above=52 mm, right=53mm, fill=orange ,draw=black] (4,0) circle (3 mm);
			
			\filldraw [above=42 mm, right=68mm, fill=red ,draw=black] (4,0) circle (3 mm);
			\filldraw [above=42 mm, right=78mm, fill=blue ,draw=black] (4,0) circle (3 mm);
			\filldraw [above=52 mm, right=68mm, fill=green ,draw=black] (4,0) circle (3 mm);
			\filldraw [above=52 mm, right=78mm, fill=orange ,draw=black] (4,0) circle (3 mm);
			
			\filldraw [above=67 mm, right=43mm, fill=red ,draw=black] (4,0) circle (3 mm);
			\filldraw [above=67 mm, right=53mm, fill=blue ,draw=black] (4,0) circle (3 mm);
			\filldraw [above=77 mm, right=43mm, fill=green ,draw=black] (4,0) circle (3 mm);
			\filldraw [above=77 mm, right=53mm, fill=orange ,draw=black] (4,0) circle (3 mm);
			
			\filldraw [above=67 mm, right=68mm, fill=red ,draw=black] (4,0) circle (3 mm);
			\filldraw [above=67 mm, right=78mm, fill=blue ,draw=black] (4,0) circle (3 mm);
			\filldraw [above=77 mm, right=68mm, fill=green ,draw=black] (4,0) circle (3 mm);
			\filldraw [above=77 mm, right=78mm, fill=orange ,draw=black] (4,0) circle (3 mm);
			
			%
			
			\draw[->, line width=3pt, draw=black ] (2.5,1.5) -- (3.8, 1.5);
			\draw[->, line width=3pt, draw=black ] (2.7,1.3) -- (2.7, 2.6);
			\node at (2.7,2.8,0.0) {\textcolor{black}{\scalebox{1.3}{$\boldsymbol{y}$}}};	
			\node at (4.,1.5,0.0) {\textcolor{black}{\scalebox{1.3}{$\boldsymbol{x}$}}};
			
			\node[above=17 mm, right=14.5mm] at (4.05,0.0,0.0) {\textcolor{white}{\scalebox{1.3}{\textbf C}}};
			\node[above=17 mm, right=24.8mm] at (4.05,0.0,0.0) {\textcolor{white}{\scalebox{1.3}{\textbf D}}};
			\node[above=27 mm, right=14.5mm] at (4.05,0.0,0.0) {\textcolor{white}{\scalebox{1.3}{\textbf A}}};
			\node[above=27 mm, right=24.8mm] at (4.05,0.0,0.0) {\textcolor{white}{\scalebox{1.3}{\textbf B}}};

			
			\draw[line width=2pt, draw=black ] (2.5,5.) -- (3.8, 5.);
			\draw[line width=2pt, draw=black ] (2.5,4.96) -- (2.5, 6.34);
			\draw[line width=2pt, draw=black ] (2.5,6.3) -- (3.8, 6.3);
			\draw[line width=2pt, draw=black ] (3.8,4.96) -- (3.8, 6.34);
			\draw[line width=2pt, draw=black ] (3.15,5.65) -- (3.82, 6.31);
			\draw[line width=2pt, draw=black ] (3.13,5.653) -- (3.82, 5.653);
			
			\node[above=-2mm, right=-3 mm] at (3.15,5.65,0.0) {\textcolor{black}{\scalebox{1.}{$\Gamma$}}};
			\node[above=-0.5mm, right=-0.5 mm] at (3.82, 5.653,0.0) {\textcolor{black}{\scalebox{1.}{$X$}}};
			\node[above=-0.5mm, right=-0.5 mm] at (3.82, 6.34,0.0) {\textcolor{black}{\scalebox{1.}{$M$}}};
			\end{tikzpicture}	
		}
		\caption{Cutout of an infinite 2D SSH lattice consisting of four sublattices. The unit cell includes one particle A, B, C, D of each sublattice. Furthermore, the high symmetry points of the first Brillouin zone are shown.} 
		\label{schema_sshgitter}
	\end{figure}
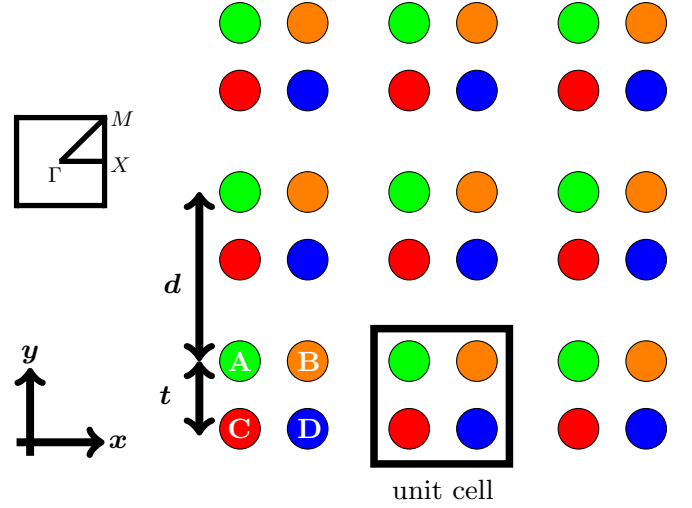
	
	For the 2D SSH lattice we now consider a configuration as shown in Fig.~\ref{schema_sshgitter}. In general it is built by SSH chains in $x$ and $y$ direction. A unit cell consists now of four particles A,B,C, and D, one of each on a 2D sublattice with lattice constant $d$. In this case, the induced dipole moment  $\mathbf{p}_{Ai}$ of NP A of the i-th unit cell is given by
	\begin{equation}
	\begin{split}
	\mathbf{p}_{Ai} &= k_0^2 \sum_{j \neq i} \alpha \mathds{G}^{\rm E}(\mathbf{r}_{Ai},\mathbf{r}_{Aj}) \mathbf{p}_{Aj} \\
	&\quad + k_0^2 \sum_{j} \alpha \mathds{G}^{\rm E}(\mathbf{r}_{Ai},\mathbf{r}_{Bj}) \mathbf{p}_{Bj} \\
	&\quad + k_0^2 \sum_{j} \alpha \mathds{G}^{\rm E}(\mathbf{r}_{Ai},\mathbf{r}_{Cj}) \mathbf{p}_{Cj} \\
	&\quad + k_0^2 \sum_{j} \alpha \mathds{G}^{\rm E}(\mathbf{r}_{Ai},\mathbf{r}_{Dj}) \mathbf{p}_{Dj}.
	\end{split}
	\label{Eq:InducedDipole2D}
	\end{equation}
	Similar equations hold for the dipole moments of the other NP B, C, and D. 
	
	As for the 1D SSH chain, we can now make a Bloch ansatz. By using the symmetry of the system we distinguish between in plane (ip) and out of plane (op) polarization of the plasmonic particles. Then we arrive at the expressions ($\nu = \rm ip, op$)
	\begin{equation}
	\mathds{M}^\nu \begin{pmatrix} p^{\nu}_A \\ p^{\nu}_B\\ p^{\nu}_C\\ p^{\nu}_D \end{pmatrix} = \frac{1}{\alpha}\begin{pmatrix} p^{\nu}_A \\ p^{\nu}_B\\ p^{\nu}_C\\ p^{\nu}_D \end{pmatrix}
	\label{Eq:EVE2D}
	\end{equation}
	for both polarization with
	\begin{equation}
	{M}^{\rm op}_{\gamma\delta} = {M}^\perp_{\gamma\delta} 
	\end{equation}
	as defined in Eq.~(\ref{mkomp}) and $\mathbf{p}^{\rm op}$ being the $z$ component of the dipole moment. For the in-plane modes we have
	\begin{eqnarray}
	\mathds{M}^{\rm ip}_{\gamma\delta} = k_0^2 \sum_{j, \gamma_i\neq \delta_j}\mathds{G}^{\rm E, ip}_{\gamma_{i}\delta_{j}}e^{\I \mathbf{k}\cdot(\mathbf{r}_{\delta_i}-\mathbf{r}_{\delta_{j}})},
	\end{eqnarray}
	with the dipole moment $\mathbf{p}^{\rm ip} = (p_x,p_y)^t$ and Green tensor
	\begin{eqnarray}
	\mathds{G}^{\rm E, ip}=\begin{pmatrix}
	G_{xx}^{\rm E} &G_{xy}^{\rm E}\\ G_{yx}^{\rm E} &G_{yy}^{\rm E}.
	\end{pmatrix}
	\label{Eq:Greenxy}
	\end{eqnarray}
	within the x-y plane.
	
\begin{figure}[h!]
		\centering
		\includegraphics[width=0.4\textwidth]{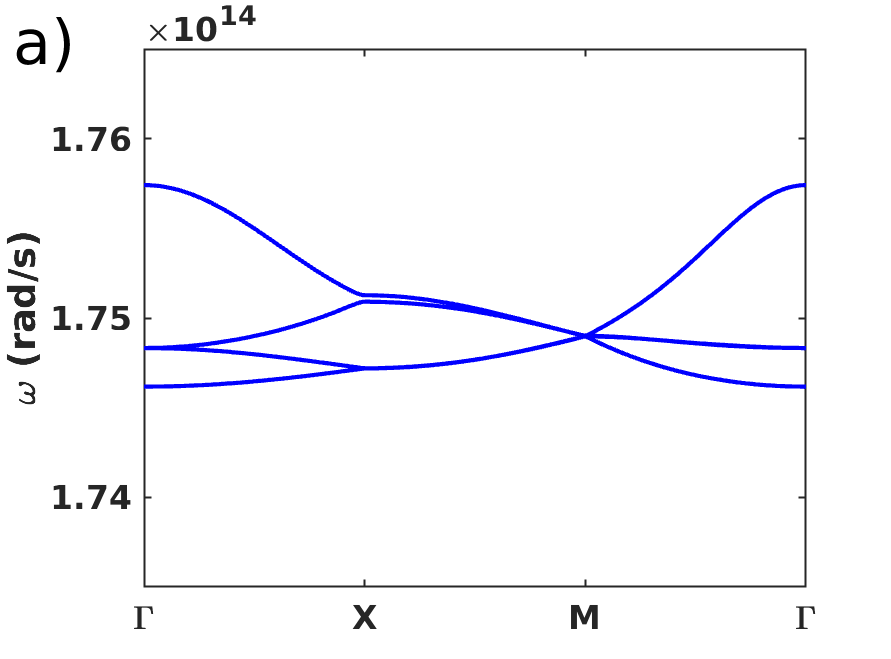}\\
		\includegraphics[width=0.4\textwidth]{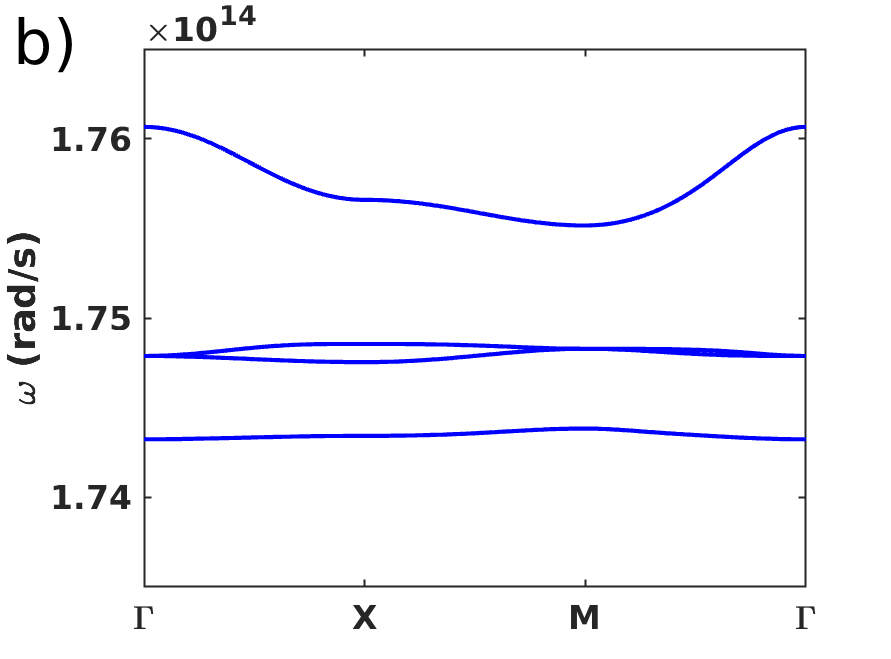}\\
		\caption{Bandstructure of infinite 2D SSH lattice for op polarization in quasi static regime for a) $\beta=1$ and b) $\beta=1.3$.}
		\label{eigenop}
	\end{figure}

\begin{figure}[h!]
		\centering
		\includegraphics[width=0.4\textwidth]{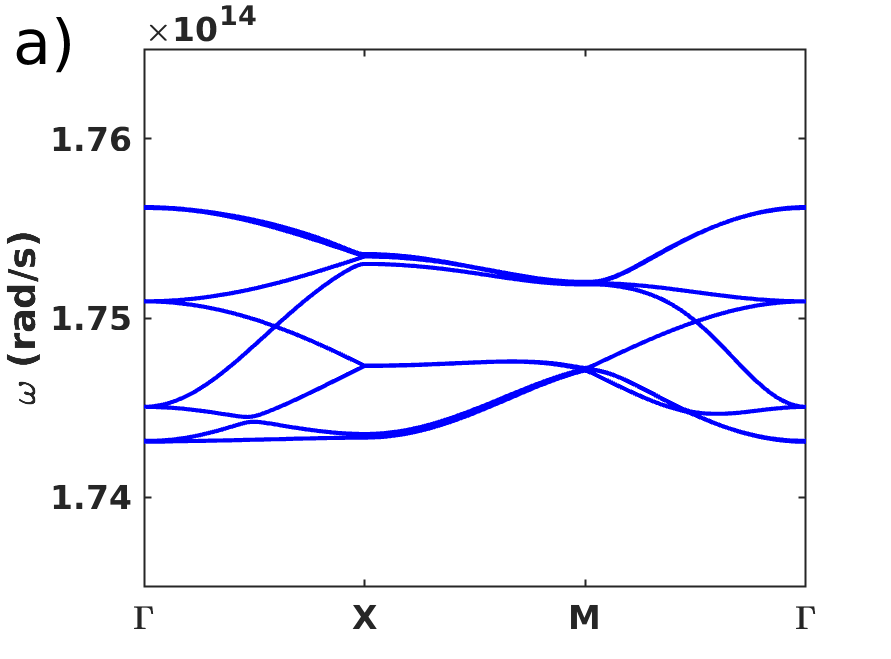}\\
		\includegraphics[width=0.4\textwidth]{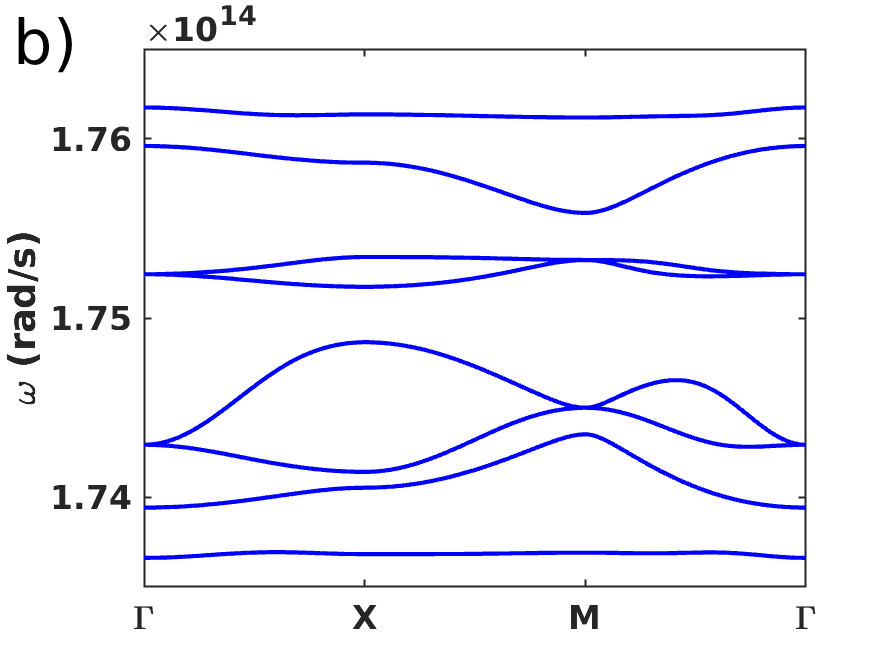}
		\caption{Bandstructure of infinite 2D SSH lattice for ip polarization in quasi static regime for a) $\beta=1$ and b) $\beta=1.3$.}	\label{eigenip}
	\end{figure}

	As for the SSH chain, the behavior of the band structure for an infinite number of particles are the same for bands with $\beta=1+x$ and $\beta=1-x$ for a given value $x$. Moreover, for $\beta=1$ the bandgaps appearing for $\beta\neq1$ close, as can be seen in Fig.~\ref{eigenop} for op polarization and in Fig.~\ref{eigenip} for ip polarization.
	
	To evaluate the topological phase of the 2D SSH lattice we determine the Zak phase~\cite{2dzakphase} by using the Wilson loop approach~\cite{quellec}.
	The Zak phase is defined by the Brillouin zone integral ($i,j \in\{x,y\}$ with $i \neq j$)
	\begin{equation}
	\vartheta_j =  d \int_{\rm BZ} \! \rd k_i \, w_j(k_i) 
	\end{equation}
	of the ``polarization''  
	\begin{equation}
	w_j(k_i) = - \frac{\ri}{2 \pi} \ln\Big(\det\Big[W_j(k_i) \Big]\Big)
	\end{equation}
	which can be evaluated by means of the Wilson loop operator
	\begin{eqnarray}
	W_j(k_i) = \prod_{n=0}^{M_j}[F_{j,\mathbf{k} + n \Delta k_j \mathbf{e}_j}]
	\end{eqnarray}
	defined via the matrix
	\begin{eqnarray}
	(F_{j,\mathbf{k}})_{mn}=\langle p_{m,\mathbf{k}}|p_{n,\mathbf{k}+\Delta k_j \mathbf{e}_j}\rangle,
	\end{eqnarray}
	where $M_j$ defines the k-space partitioning by the relation $(M_j +1)\Delta k_j=\frac{2\pi}{d}$, and $m,n\in\{1,...,\tilde{N}\}$. Here $\tilde{N}$ is the number of the bands below the considered band gap and therefore defines the rank of the matrix $F_{j,\mathbf{k}}$. The vectors $|p_{n,\mathbf{k}+\Delta k_j\mathbf{e}_j}\rangle$ are given by the eigenvector solutions $(p_A^\nu, p_B^\nu, p_C^\nu, p_D^\nu)^t$ of the eigenvalue equation (\ref{Eq:EVE2D}) for the corresponding $\mathbf{k}+\Delta k_j\mathbf{e}_j$ and $M^\nu$ restricted only to $\nu = x$, $\nu = y$, or $\nu = z$ subspace. 
	
\begin{figure}[h!]
		\centering
		\includegraphics[width=0.4\textwidth]{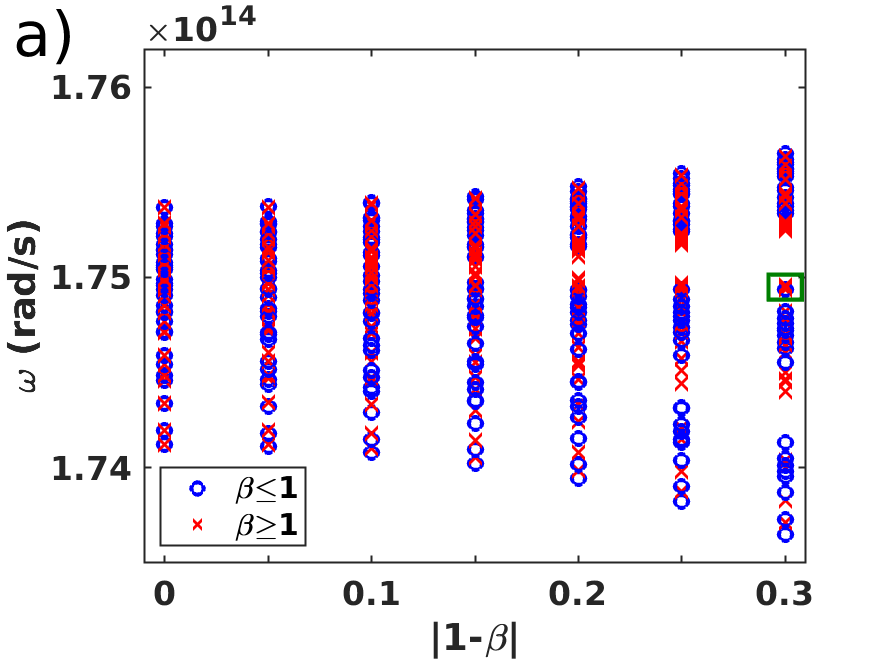}\\
		\includegraphics[width=0.4\textwidth]{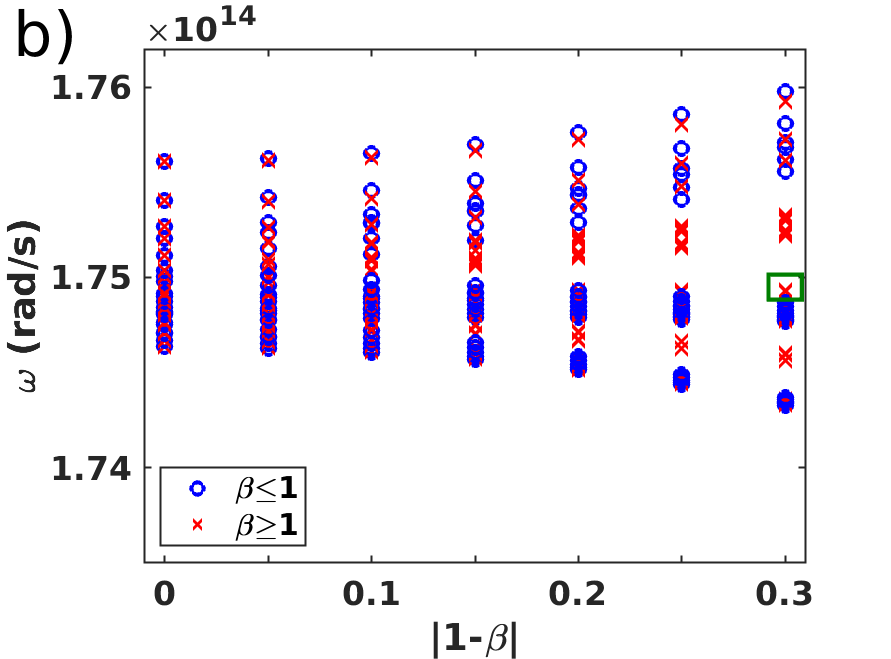}\\
		\caption{For a finite 2D SSH lattice of 36 particles real parts of the complex eigenfrequencies of the lattice modes determined by Eq.~(\ref{Eq:InducedDipole}) are shown as a function of $\beta$ for a) ip and b) op polarization. For $\beta < 1$ (blue crosses) the Zak phase is $\vartheta_{x/y} = 0$ for all band gaps and for $\beta > 1$ (red crosses) the Zak phase is $\vartheta_{x/y} = \pi$ for the first and third band gap. The green box indicates the corner modes at $\beta = 1.3$, the other two red crosses in the band gaps are the edge modes.}
		\label{omegabetaeigen}
	\end{figure}

\begin{figure}[h!]
		\centering
		\includegraphics[width=0.4\textwidth]{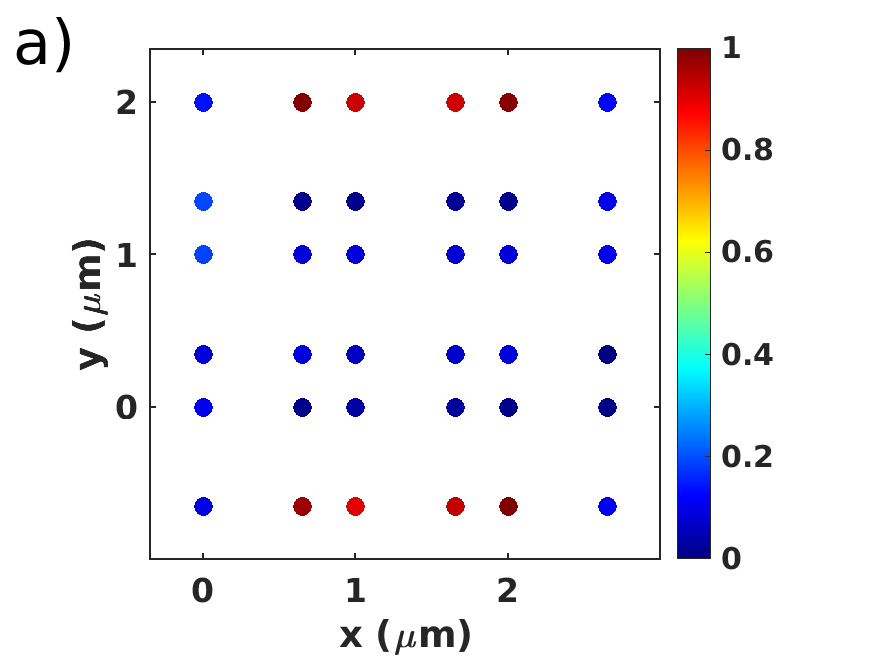}\\
		\includegraphics[width=0.4\textwidth]{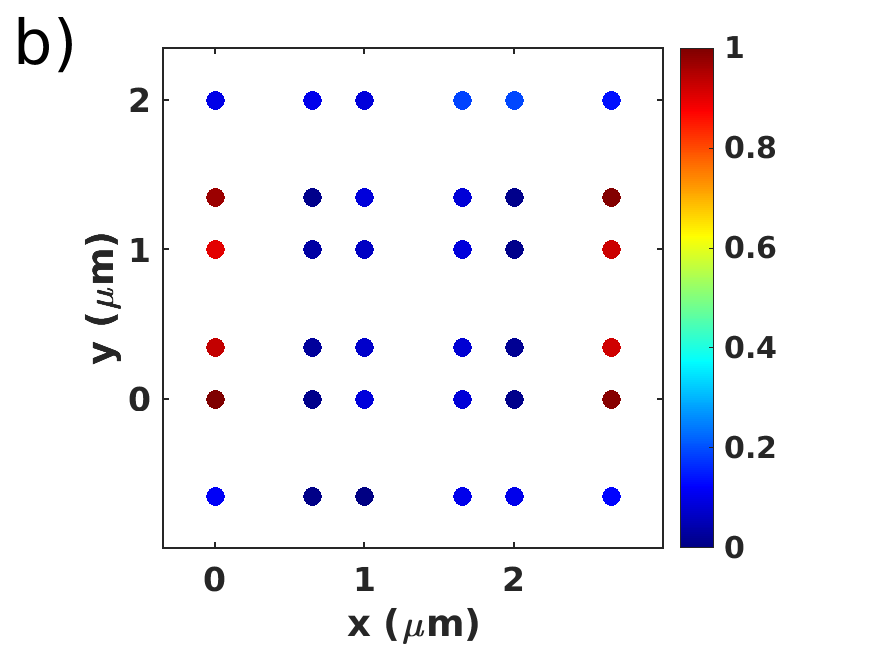}\\
                \includegraphics[width=0.4\textwidth]{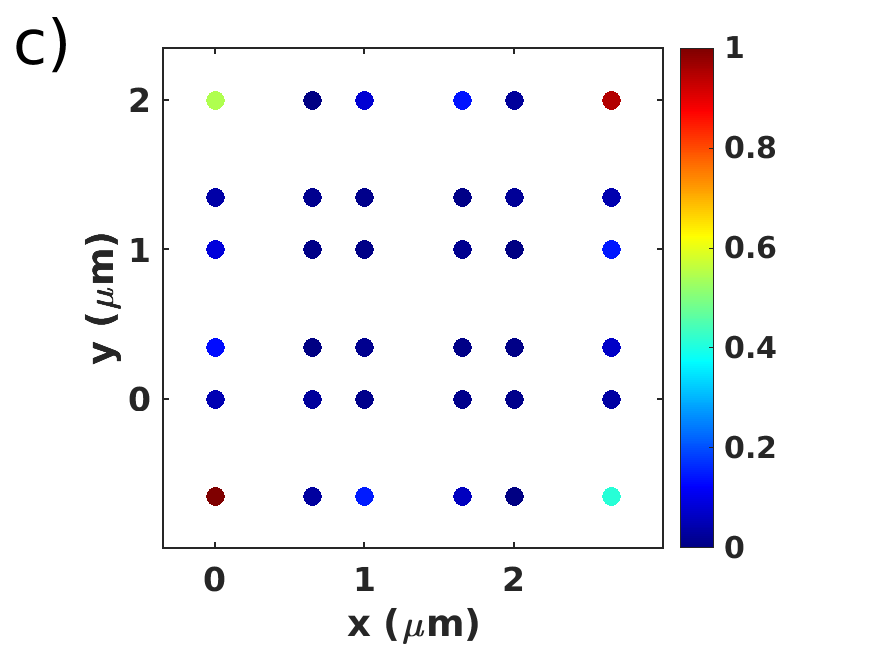}
		\caption{Normalized absolute values of the dipole moments $|p_{A,B,C,D}^{\rm op}|$ of two op edge modes at frequency $1.746\times10^{14}\,{\rm rad/s}$ when exciting a particle in the edge at the a) bottom or b) left and c) a corner mode at the frequency $1.749\times10^{14}\,{\rm rad/s}$ when exciting the particle at the left bottom corner for $\beta = 1.3$ as found in Fig.~\ref{omegabetaeigen}.}
		\label{peigen}
	\end{figure}

	For the polarization of the particles in $x,y,z$ direction, we obtain the same results as in Ref.~\cite{sshgitter}, namely for the polarization in $x,y$ in the first and third band gap we have $(\vartheta_x,\vartheta_y)=(\pi,\pi)$ and in the second and fourth band gap we find  $(\vartheta_x,\vartheta_y)=(0,0)$ if $\beta > 1$. For polarization in $z$ for the first band and second gap we get $(\vartheta_x,\vartheta_y)=(\pi,\pi)$ for $\beta>1$ and $(\vartheta_x,\vartheta_y)=(0,0)$ for $\beta<1$. Following to the bulk-edge correspondence there will be topological edge modes in the first and third band gap  in the topological non-trivial phase if the lattice is finite in either x- or y-direction. If the lattice is finite in $x$ and $y$ direction then we also expect corner states appearing at the corners of the lattice~\cite{sshgitter}. To illustrate the topological edge and corner modes we show the band structures for ip and op polarization in Fig.~\ref{omegabetaeigen}. It can be seen that there are additional modes in the band gap for $\beta>1$ compared to $\beta<1$ for a finite system. These modes are localized at the edges or corners of the lattice structure. This localization will be evident later when we determine the energy density in close vicinity of the SSH lattice. Of course, this localization can also be confirmed by plotting the eigenvector solutions for the dipole moments of the lattice. In Fig.~\ref{peigen} we show the normalized values of the eigensolutions of the dipole moment for two edge and the corner mode for $\beta = 1.3$ and op polarization. Similar results can be obtained for the ip polarization.
	
	\section{Numerical results}
	
	In the following we discuss the numerical results of the spectral energy density $u_\omega = u_1+u_2+u_3$ for the edge and corner modes in 1D SSH chain and 2D SSH lattice. To this end, we assume that all NPs have a temperature of 350 K whereas the background has a lower temperature of $T_b=300\,{\rm K}$. 
	
	\begin{figure}[h!]
		\centering
		\includegraphics[width=0.4\textwidth]{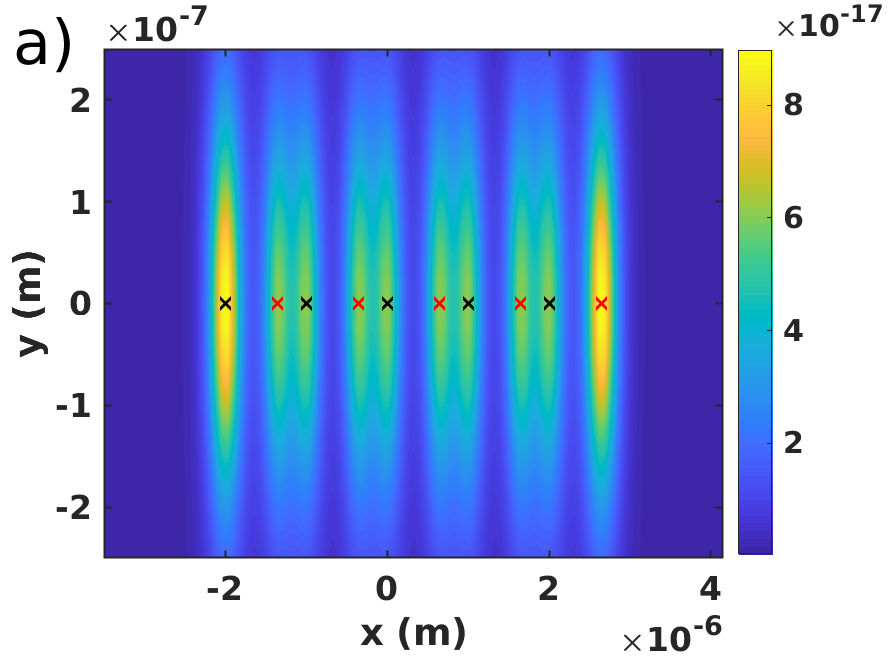}\\
		\includegraphics[width=0.4\textwidth]{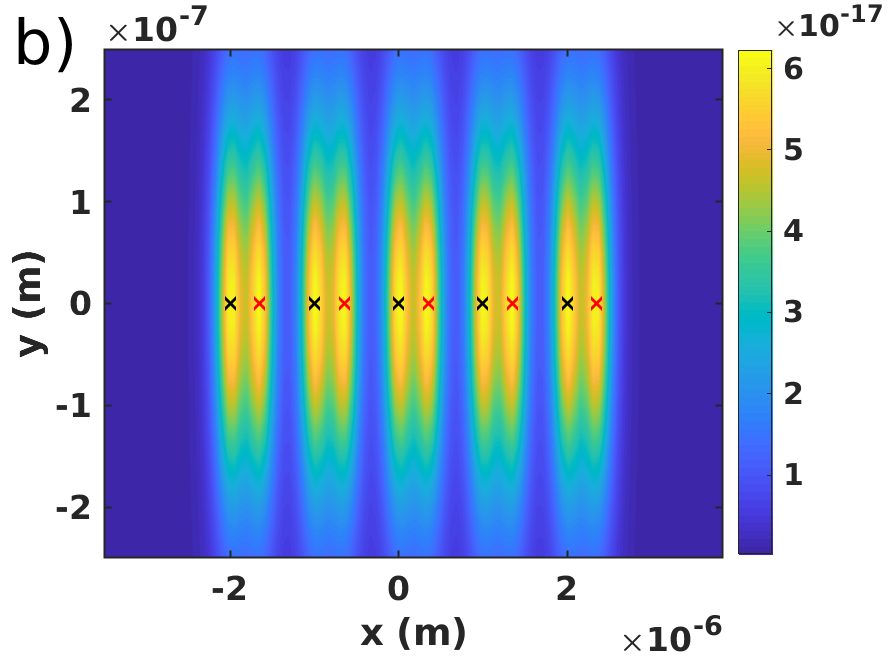}
		\caption{Full spectral energy density $u_\omega$ in (Js/m$^3$) (longitudinal and transversal polarization) above an SSH chain of NPs for a) $\beta=1.3$ and b) $\beta=0.7$. Here we choose 10 particles with radius $R=100\,{\rm nm}$. The spectral energy density is calculated 300 nm above the NPs at the edge mode frequency $\omega = 1.7493\cdot10^{14}$ rad/s for $\beta=1.3$. Black (A) and red (B) crosses mark the positions of the NPs of both sublattices.}
		\label{betavgl}
	\end{figure}
	
	\begin{figure}[h!]
		\centering
		\includegraphics[width=0.4\textwidth]{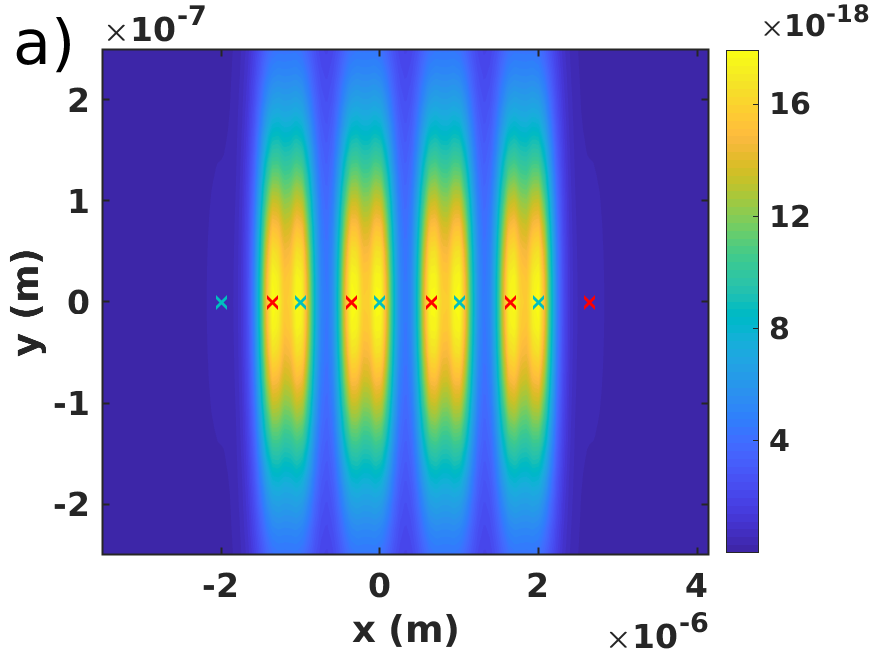}\\
		\includegraphics[width=0.4\textwidth]{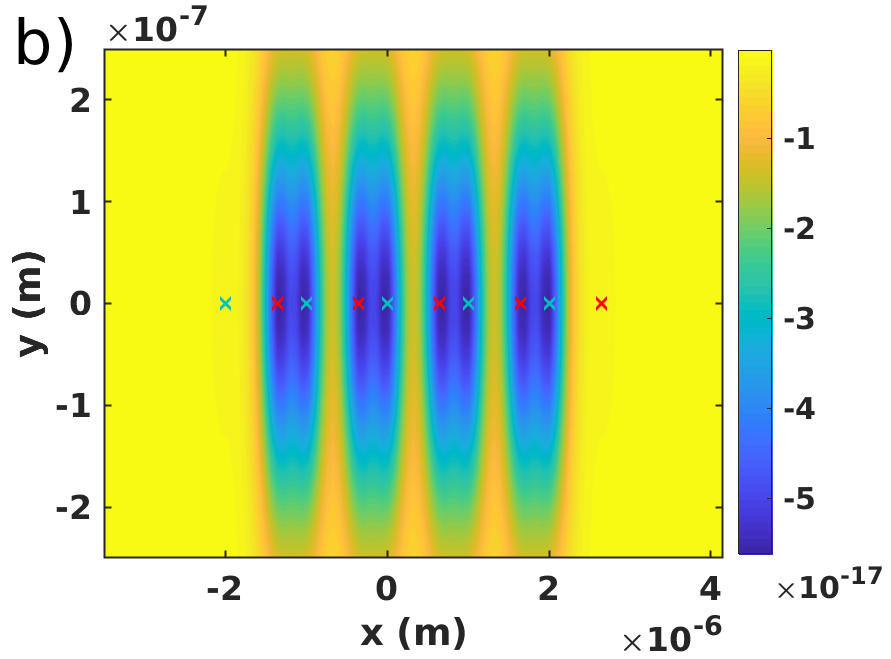}\\
		\includegraphics[width=0.4\textwidth]{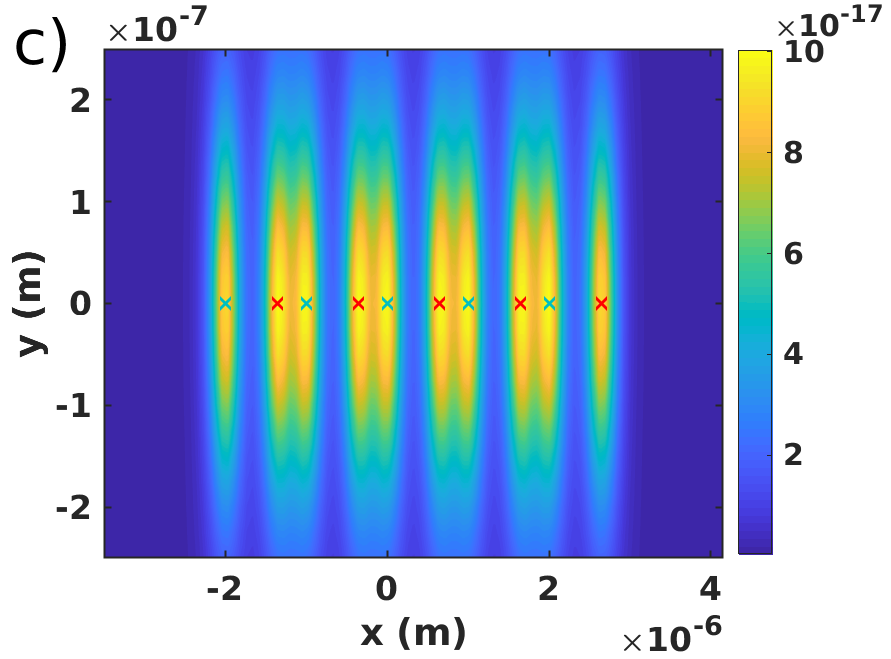}\\
		\caption{Spectral energy density terms a) $u_1$, b) $u_2$, and c) $u_3$ in (Js/m$^3$) for $\beta=1.3$. The other parameters are the same as in Fig.~\ref{betavgl}.}
		\label{terme_ohnepol}
	\end{figure}
	
	\begin{figure}[h]
		\centering
		\includegraphics[width=0.4\textwidth]{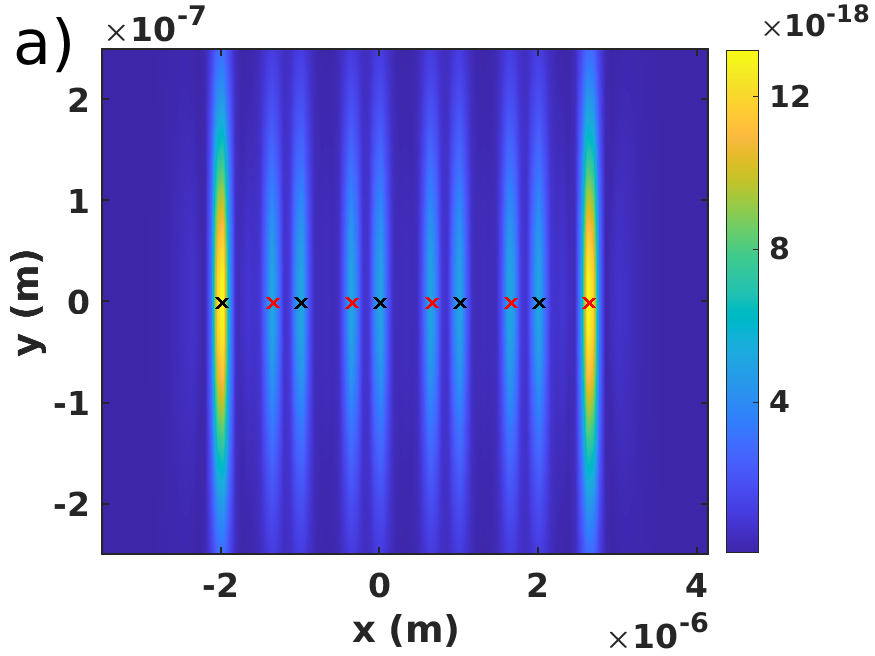}\\
		\includegraphics[width=0.4\textwidth]{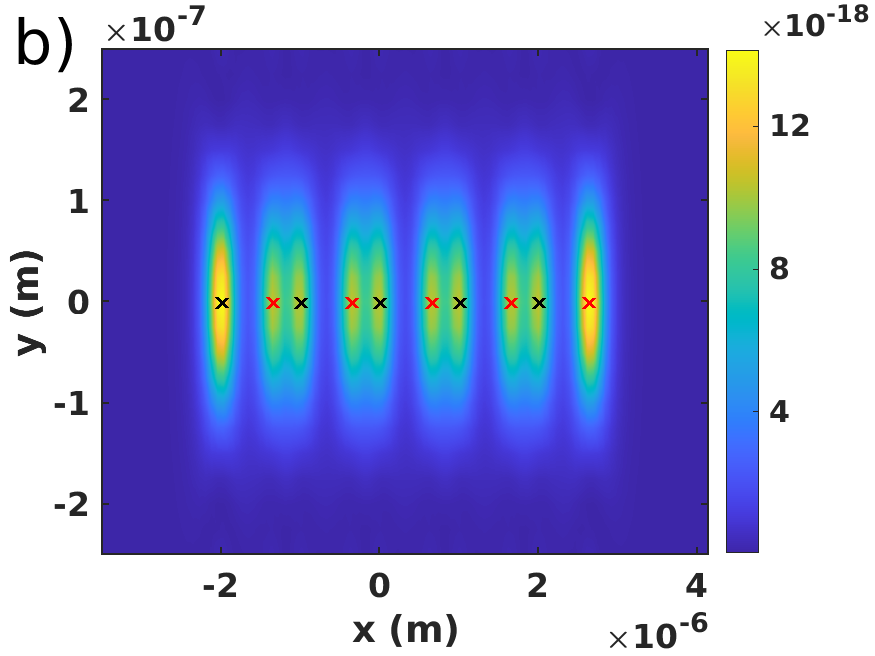}
		\caption{Spectral energy density for a) longitudinal and b) transversal polarization in (Js/m$^3$). The parameters are the same as in Fig.~\ref{betavgl}. }
		\label{longtrans}
	\end{figure}

	\subsection{SSH chain}
	
	In Fig. \ref{betavgl} we plot the full spectral energy density for $\beta>1$ at frequency $\omega = 1.7493\cdot10^{14}$ rad/s, i.e.\ the contribution of longitudinal and transversal modes together. It can be seen that the edge modes appearing at that frequency dominate the spectral energy density at the edges of the chain. In contrast, for $\beta<1$ at the same frequency the energy density is the same above each particle of the chain. This means that the topological protected edge modes can in principle be measured with a near-field microscope like the  TINS, TRSTM, or the SNoiM. 
	To get a better insight of the contribution of the terms $u_1$, $u_2$, and $u_3$ we show the spectral energy density at the edge mode frequency in the topological non-trivial case for all three terms separately in Fig.~\ref{terme_ohnepol}. From these plots it becomes clear the behavior of $u_2$, which is linear in $\tildeblockt$, is responsible for the dominating edge mode contributions, whereas $u_1$ and $u_2$ have smaller energy density contributions at the edges than in the inner part of the chain.

	Now, let us turn to the contributions of the longitudinal and transversal modes. It can be expected from the fact that for both polarizations there is an edge mode at the same frequency as shown in Fig.~\ref{modes}. This is confirmed by the longitudinal and transversal spectral energy density above the chain of NP shown in Fig.~\ref{longtrans}. In both cases the general behavior is the same, i.e.\ the edge modes give a dominant energy density contribution at the edges of the chain. {Here, this enhanced energy density at the edge is more pronounced for the longitudinal polarization than for the transversal polarization.}
	
	\subsection{2D SSH lattice}
	
\begin{figure}[h]
		\centering
		\includegraphics[width=0.4\textwidth]{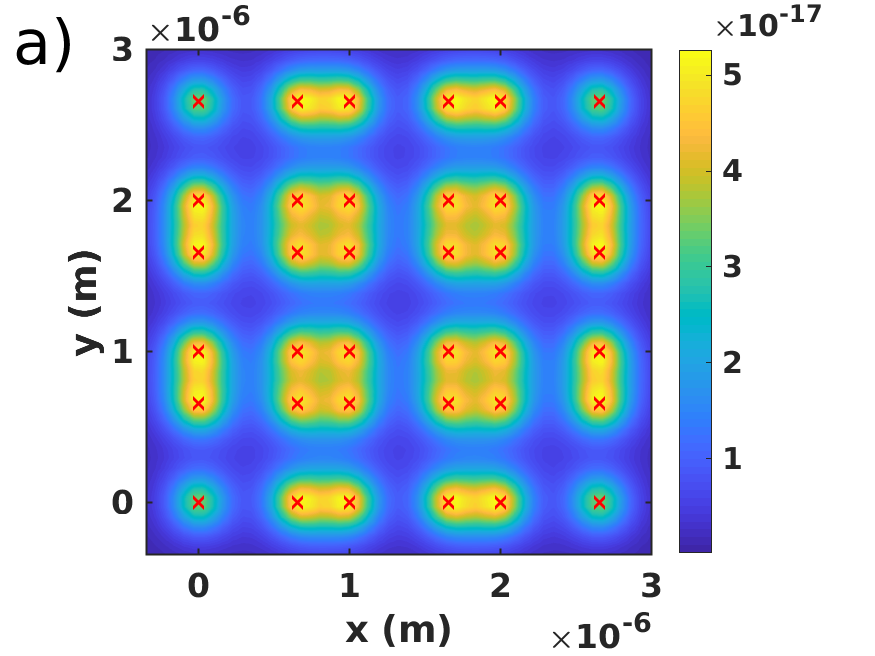}\\
		\includegraphics[width=0.4\textwidth]{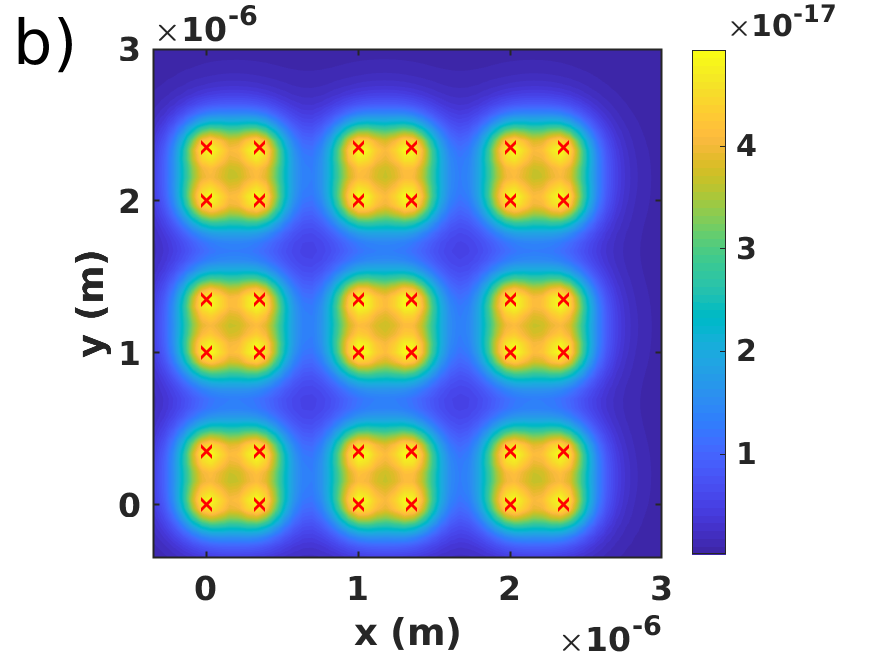}
		\caption{Full spectral energy density in (Js/m$^3$) of a 2D SSH lattice for a) $\beta=1.3$ and b) $\beta=0.7$ for a lattice of 36 NPs (positions are marked via red crosses) at the edge mode eigenfrequency $\omega = 1.75592\cdot10^{14}$ rad/s. All other parameters are the same as in Fig.~\ref{betavgl}.}
		\label{2Dbetarm}
	\end{figure}

\begin{figure}[h]
		\centering
		\includegraphics[width=0.4\textwidth]{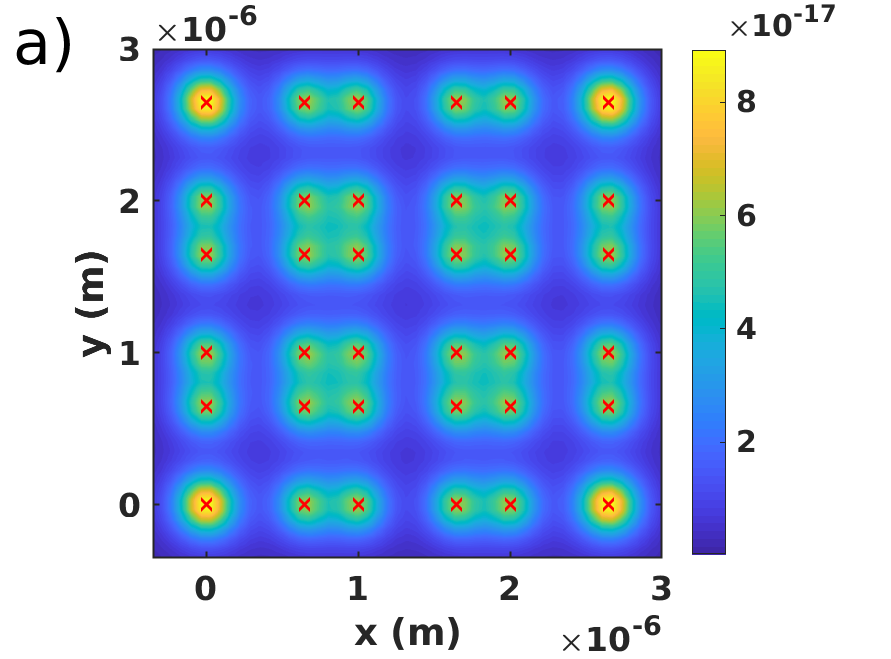}\\
		\includegraphics[width=0.4\textwidth]{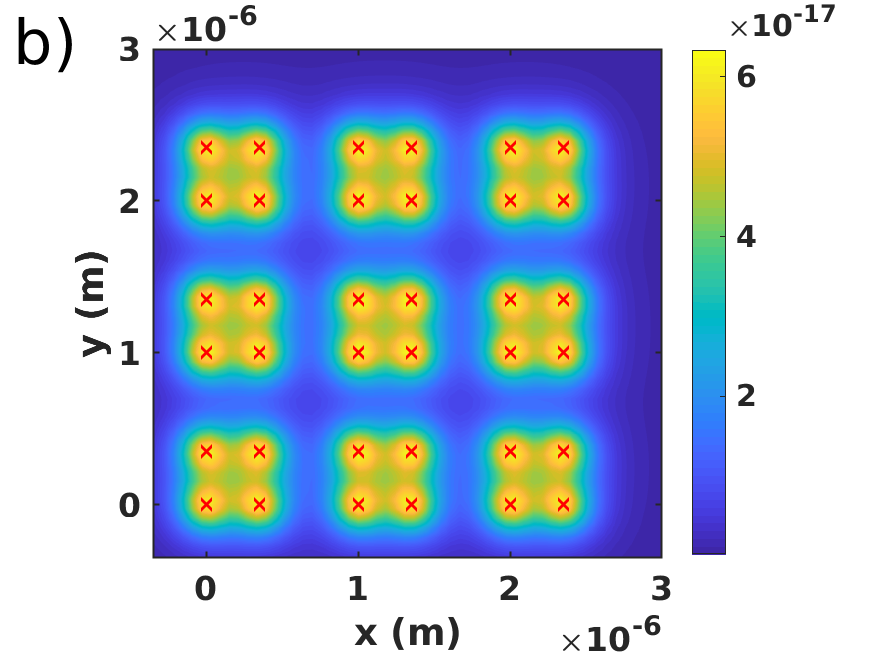}
		\caption{Full spectral energy density in (Js/m$^3$) of a 2D SSH lattice for a) $\beta=1.3$ and b) $\beta=0.7$ for a lattice of 36 NPs (positions are marked via red crosses) at the corner mode eigenfrequency $\omega = 1.7494\cdot10^{14}$ rad/s. All other parameters are the same as in Fig.~\ref{betavgl}.}
		\label{2Dbetaem}
	\end{figure}

	Let's turn to the investigation of the spectral energy above the 2D SSH lattice. {In general, it shows the same behavior as for the chain, therefore, only full spectral energy density without a detailed discussion of the separate terms and polarizations will be exemplarily considered.} As can be seen in Fig.~\ref{2Dbetarm} and Fig.~\ref{2Dbetaem} for a frequency corresponding to the eigenfrequency of the edge or corner mode the energy density is enhanced at the edges or corners, respectively, when choosing the non-trivial topological phase. Again we can conclude that the edge and corner modes are due to this enhancement accessible by scattering type near-field probes like TINS, TRSTM, and SNoiM. 
	
	\subsection{Stability}
	
\begin{figure}[h!]
		\centering
		\includegraphics[width=0.4\textwidth]{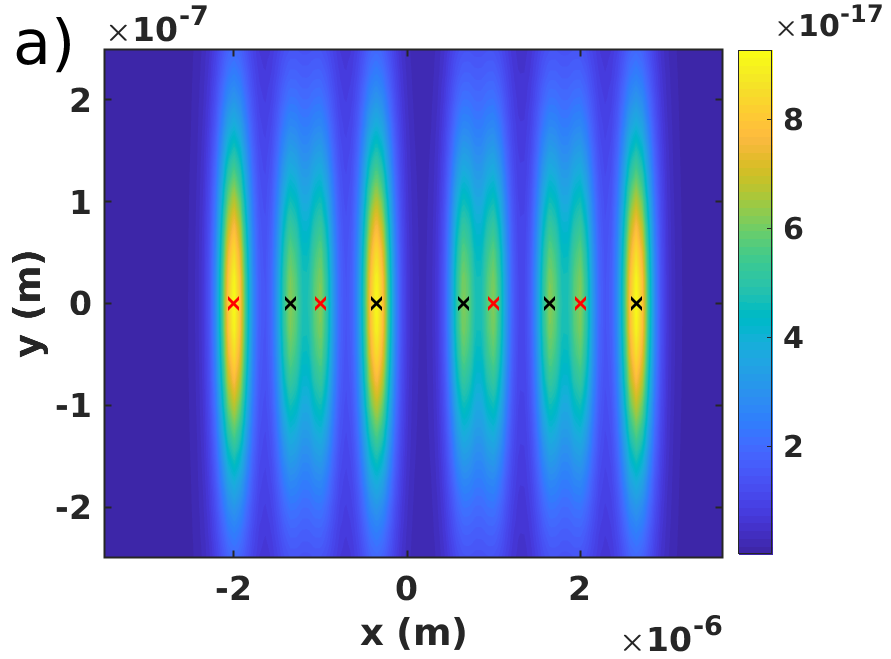}\\
		\includegraphics[width=0.4\textwidth]{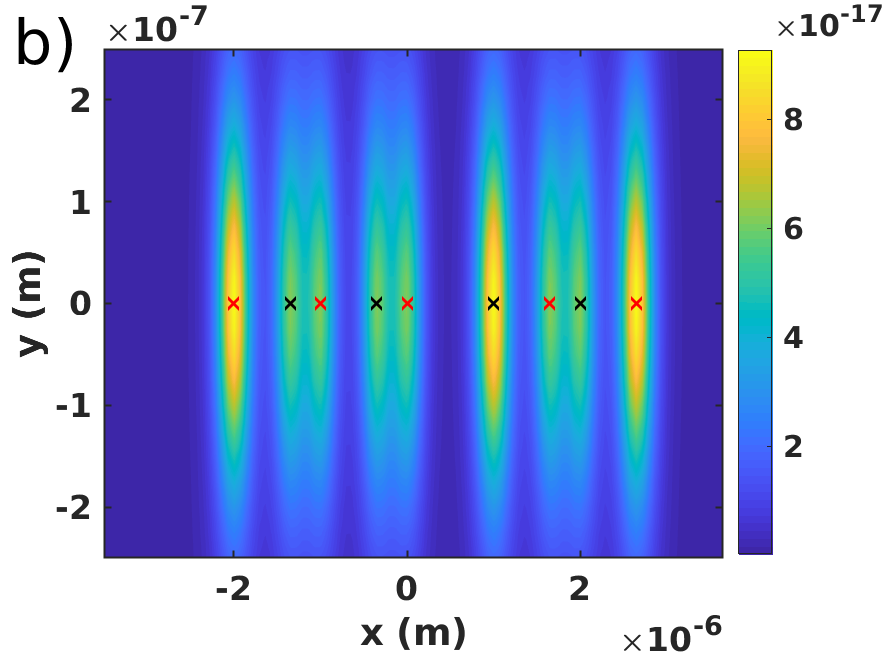}
		\caption{Full spectral energy density in (Js/m$^3$) for $\beta=1.3$ for a chain of 10 NPs at the a) edge mode frequency with {a vacancy in the NP chain} of the A (top) and B (bottom) sublattice. All other parameters are the same as in Fig.~\ref{betavgl}.}
		\label{defekt}
	\end{figure}

	As topological generated edge and corner modes are robust we show how this translates for the near-field energy density. That these modes are robust under small perturbations for example in the positioning of the NPs is known~\cite{2dzakphase}. Here we show that the topological edge modes are even robust to defects within the lattice which are not treatable within perturbation theory~\cite{2dzakphase}. Furthermore, due to such defects new edges and therefore new edge modes can appear. Therefore we include defects in the chain or lattice including positions in the chain or lattice where the NP is absent. In Fig.~\ref{defekt} there are two examples for the SSH chain by removing either an "A" or "B" NP in the chain. It can be seen that the edge modes of the previously defectless chain are still enhancing the energy density at the edges. Furthermore, due to the fact that by removing an NP we have now two separate SSH chains, we find also edge modes in the produced gap for the subchain which ends by an "A" and "B" NPs. The subchain which has either two "A" or two "B" NPs at its ends only supports one edge mode as can directly be seen in the near-field energy density.
	
\begin{figure}[h!]
		\centering
		\includegraphics[width=0.4\textwidth]{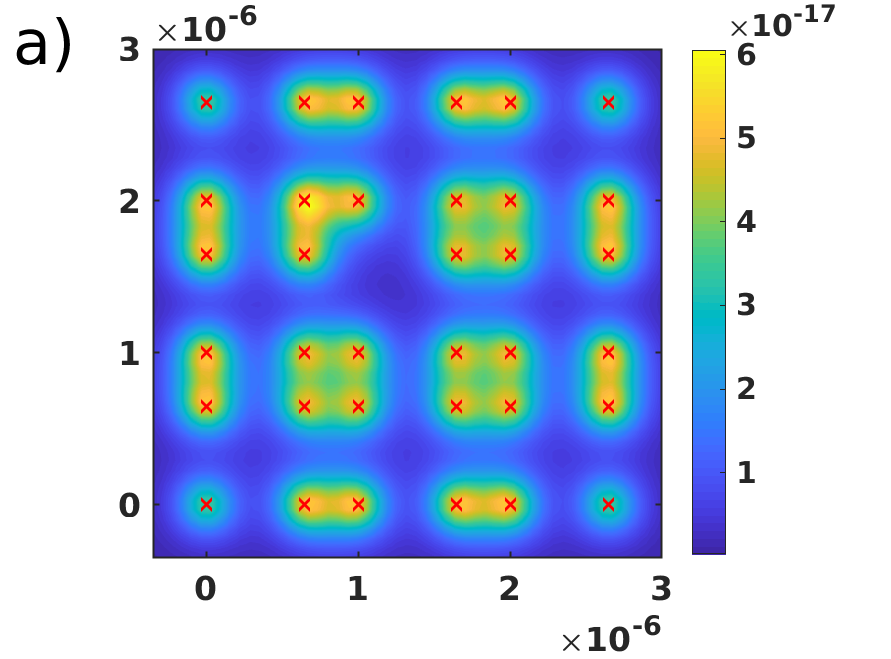}\\
		\includegraphics[width=0.4\textwidth]{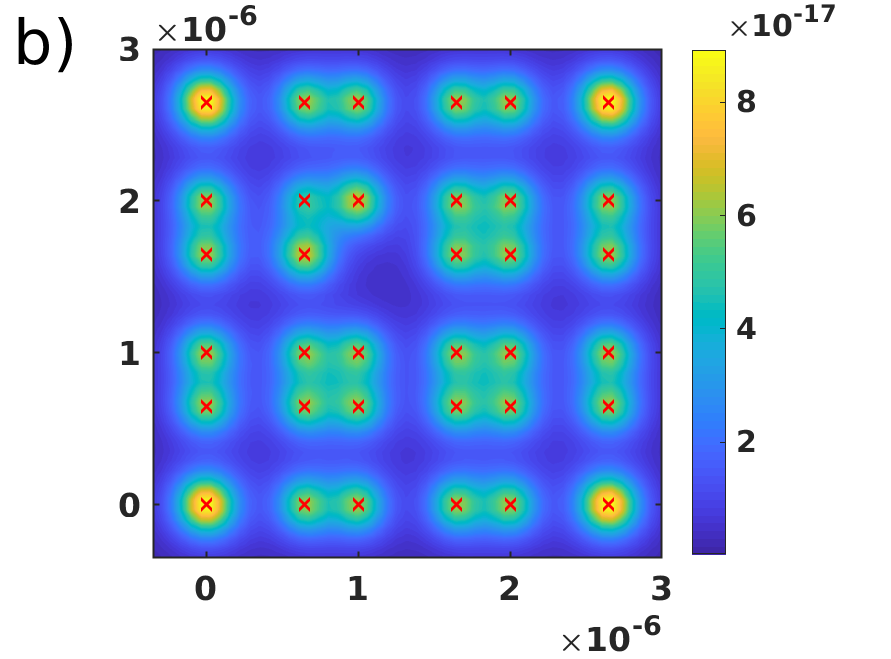}\\
		\caption{Full spectral energy density in (Js/m$^3$) for $\beta=1.3$ for a lattice of 36 particles (positions are marked via red crosses) at the a) edge mode frequency $\omega=1.7559\cdot10^{14}$ rad/s and b) at the corner mode frequency $\omega=1.74942\cdot10^{14}$ rad/s with an absent NP in the middle of the chain. All other parameters are the same as in Fig.~\ref{betavgl}.}
		\label{udefekt}
	\end{figure}

\begin{figure}[h!]
	\centering
	\includegraphics[width=0.4\textwidth]{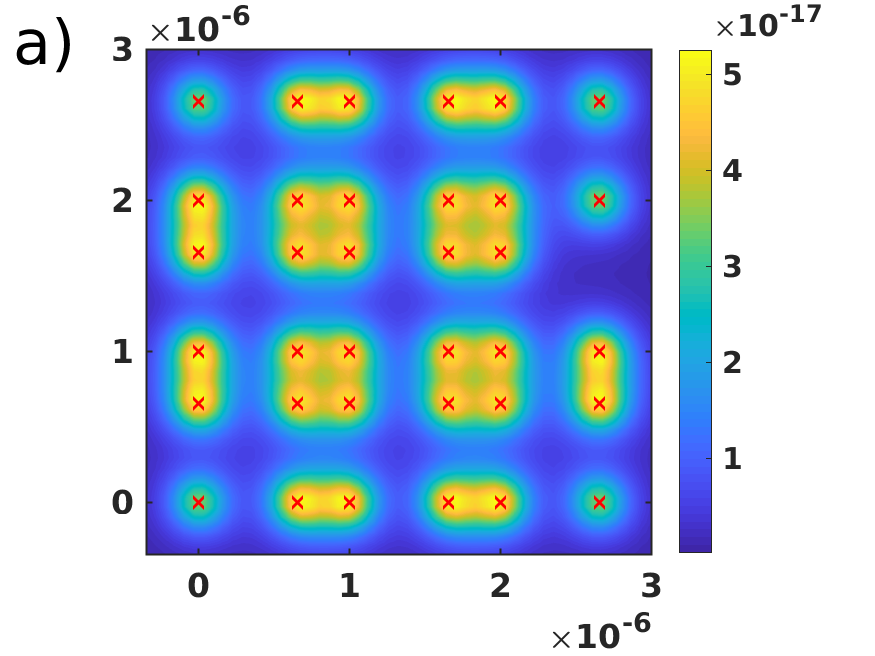}\\
	\includegraphics[width=0.4\textwidth]{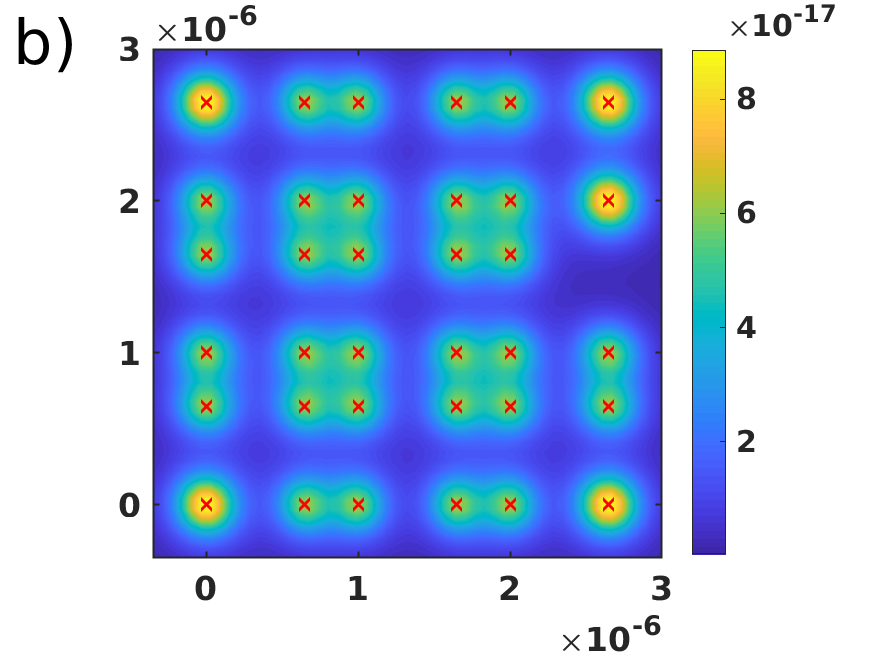}\\
	\caption{Full spectral energy density in (Js/m$^3$) for $\beta=1.3$ for a lattice of 36 particles (positions are marked via red crosses) at a) the edge mode frequency $\omega=1.7559\cdot10^{14}$ rad/s and at the b) corner mode frequency $\omega=1.7494\cdot10^{14}$ rad/s. All other parameters are the same as in Fig.~\ref{betavgl}.}
	\label{udefekt2}
\end{figure}

	If we add defects to the lattice by removing an NP in the bulk part of the lattice than the edge and corner modes are basically unaffected as can be seen in Fig.~\ref{udefekt} even though the energy density close to the defect can be affected as can be nicely seen in Fig.~\ref{udefekt} a). Now, when adding a defect at the edges at the lattice as shown in Fig.~\ref{udefekt2} then the edge and corner modes are basically unaffected in the region far away from the defect. However in the direct neighborhood of the defect in our case a new corner mode appears in Fig.~\ref{udefekt2} a) and the edge mode is partially "destroyed" in Fig.~\ref{udefekt2} b). Therefore we can conclude that defects whicht might be produced by nanofabricating SSH chains and lattices are not at all affecting the edge or corner modes if the defects are in the middle part of the chain or lattice.

	\section{Conclusion}
	
	We have derived a generalized expression for the near-field energy density for $N$ dipoles each thermalized at its own local temperature. In contrast to former expressions ours includes the possibility of an arbitrary background which can have a different temperature than {that of} the dipoles. Furthermore, we also include the magnetic part of the energy density which also allows us to relate it to the LDOS. We have applied this
	expression to determine the near-field energy density of a 1D SSH chain and a 2D SSH lattice of InSb plasmonic NPs
	showing that the edge and corner modes enhance the energy density {in the vicinity} of the edges and corners of such structures within
	the topological non-trivial phase. {Of course, similar results can be obtained with other plasmonic or phonon-polaritonic materials like SiC, GaN, doped Si, etc. which have a plasmon or phonon-polariton resonance in the infrared.} 
	Apart from its fundamental character this observation makes clear that the edge and
	corner modes are in principle {experimentally} accessible with near-field thermal profilers such as the TINS, TRSTM, and SNoiM which
	opens a new perspective to studying these topological states and the topological phase transition. Finally, we have
	also shown that defects due to possible fabrication errors will not affect the edge and corner modes as long as the
	defects do not sit at the edges or corners of interest.

	\acknowledgments
	\noindent
	
	S.-A.\ B.\ acknowledges support from Heisenberg Programme of the Deutsche Forschungsgemeinschaft (DFG, German Research Foundation) under the project No. 404073166. Z.A. is supported by NSFC (No. 12027805), and Shanghai Science and Technology Committee grant (No. 18JC1420402). The authors further acknowledge support from the Sino-German Center for Research Promotion (No. M-0174).
	
	\appendix

	\section{Proof of Eq.~(\ref{eqrelation})} \label{relation}

	Here, we proof or derive Eq.~(\ref{eqrelation}). To this end, we calculate the general correlation function
	\begin{eqnarray}
\langle\mathbf{E}_i\otimes\mathbf{E}_j\rangle
		\label{ee}
	\end{eqnarray}
        in two different equivalent ways and compare the results which will allow us to obtain Eq.~(\ref{eqrelation}). Note that here
	we use the short-hand notation $\mathbf{E}_i=\mathbf{E}(\mathbf{r}_i)$ which is the electric field at the position of 
        the $i$-th particle and correspondingly for the index $j$. Since the electric field is given by the sum of the fields generated 
        by the particles and the part generated by the background field, it can be expressed as~\cite{nteilchen}
	\begin{eqnarray}
	\begin{pmatrix} \mathbf{E}_{1} \\ \vdots \\ \mathbf{E}_{N} \end{pmatrix} =\boldsymbol{DT}^{-1}\begin{pmatrix} \mathbf{p}_1^{\,\, fl} \\ \vdots \\ \mathbf{p}_N^{\,\, fl}\end{pmatrix}+ (\boldsymbol{1}+\boldsymbol{DT}^{-1}\boldsymbol{A})\begin{pmatrix} \mathbf{E}_1^{b} \\ \vdots \\ \mathbf{E}_N^{\rm b}\end{pmatrix}.
	\label{eqe}
	\end{eqnarray}
  	where we introduce
 	\begin{eqnarray}
           \boldsymbol{1}_{ij} &=& \delta_{ij}\mathds{1}, \\
	   \boldsymbol{A}_{ij} &=& \epsilon_0\delta_{ij}\alpha,\\
		\boldsymbol{D}_{ij} &=& \frac{k_0^2}{\epsilon_0}\mathds{G}_{ij}^{\rm E}.
	\end{eqnarray}
        So, Eq.~(\ref{eqe}) is just another formulation of Eq. (\ref{eqg}). Inserting this ansatz into the correlation function in Eq.~(\ref{ee}) we obtain
\begin{equation}
  \begin{split}
  \langle \mathbf{E}_{i}\otimes \mathbf{E}_{j}\rangle&=\sum_{k}\Bigg\{\Big(\boldsymbol{D}\blockt\Big)_{ik}b(T)\chi\Big(\boldsymbol{D}\blockt\Big)_{jk}^\dagger \notag \\
	  &\quad+\Big(\boldsymbol{D}\blockt\boldsymbol{A}\Big)_{ik}a(T)\frac{\mathds{G}_{kj}^{\rm E}-{\mathds{G}_{jk}^{\rm E}}^\dagger}{2\ri} \notag \\
	  &\quad+a(T)\frac{\mathds{G}_{ik}^{\rm E}-{\mathds{G}_{ki}^{\rm E}}^\dagger}{2\ri}\Big(\boldsymbol{D}\blockt\boldsymbol{A}\Big)_{jk}^\dagger\notag \\
	  &{\small\quad+\sum_l a(T)\Big(\boldsymbol{D}\blockt\boldsymbol{A}\Big)_{ik}\frac{\mathds{G}_{kl}^{\rm E}-{\mathds{G}_{lk}^{\rm E}}^\dagger}{2\ri}\Big(\boldsymbol{D}\blockt\boldsymbol{A}\Big)_{jl}^\dagger\Bigg\} } \notag \\
	  &\quad+a(T)\frac{\mathds{G}_{ij}^{\rm E}-{\mathds{G}_{ji}^{\rm E}}^\dagger}{2\ri}. \label{gl1}
\end{split}
\end{equation}
where the {indices of the sums are runninv} over all particles. In addition, here we used the fluctuation-dissipation theorem for the dipolar moments of the isotropic NPs~\cite{nteilchen}
\begin{equation}
  \langle \mathbf{p}^{fl^*}_i \otimes \mathbf{p}^{fl}_j \rangle = \delta_{ij}b(T)\chi\mathds{1}
\end{equation}
and fields ~\cite{Agarwal,Eckhardt}
\begin{equation}
	\langle \mathbf{E}^{b}_i \otimes \mathbf{E}^{b^*}_j \rangle =  a(T) \frac{\mathds{G}_{ij}^{\rm E}-{\mathds{G}_{ji}^{\rm E}}^{\dagger}}{2\I}
\end{equation}
with $a(T)=2 \omega \mu_0\Theta$ and $b(T)=\frac{2\epsilon_0}{\omega}\Theta$.
These are just the same expressions as used in Eq.~(\ref{Eq:FDTp}) and Eq.~(\ref{Eq:FDTF}) with $T_i = T_b = T$. Hence we are assuming global
thermal equilibrium at temperature $T$.

On the other hand, we can directly evaluate the correlation function in Eq.~(\ref{ee}) by first evaluating the total field in presence of the scatterers by assuming that we have some current density $\mathbf{j}$ at position $\mathbf{r}'$ which generates a field 
\begin{equation}
	\mathbf{E}^0(\mathbf{r}) = \ri \omega \mu_0 \mathds{G}^{\rm E} (\mathbf{r},\mathbf{r}') \mathbf{j}.
\label{Eq:E0}
\end{equation}
Then the total field at the position of the dipole $i$ is
\begin{eqnarray}
	\mathbf{E}_i = \sum_k(\boldsymbol{1} +\boldsymbol{D}\blockt\boldsymbol{A})_{ik}\mathbf{E}_k^0
\end{eqnarray}
which is basically Eq.~(\ref{eqe}) without the contribution of the fluctuational sources. By inserting Eq.~(\ref{Eq:E0}) we obtain
\begin{eqnarray}
	\mathbf{E}_i = \ri \omega \mu_0 \mathds{G}^{\rm full}(\mathbf{r}_i,\mathbf{r}') \mathbf{j}
\end{eqnarray}
with
\begin{equation}
	\mathds{G}^{\rm full}(\mathbf{r}_i,\mathbf{r}')=\mathds{G}^{\rm E}(\mathbf{r}_i,\mathbf{r}')+\sum_k \big(\boldsymbol{D}\blockt\boldsymbol{A}\big)_{ik}\mathds{G}^{\rm E} (\mathbf{r}_k,\mathbf{r}').
\end{equation}
Since we are in global equilibrium, the correlation function in Eq.~(\ref{ee}) is determined by the fluctuation-dissipation theorem as
\begin{equation}
  \langle \mathbf{E}_{i}\otimes \mathbf{E}_{j}\rangle = a(T)\frac{\mathds{G}_{ij}^{\rm full}-\mathds{G}_{ji}^{\rm full\dagger}}{2\ri} 
\end{equation}
so that by inserting the full Green function we obtain
\begin{equation}
\begin{split}
	\langle \mathbf{E}_{i}\otimes \mathbf{E}_{j}\rangle &= a(T)\Bigg[\sum_k\Bigg\{\frac{\Big(\boldsymbol{D}\blockt\boldsymbol{A}\Big)_{ik}\mathds{G}_{kj}^{\rm E}}{2\ri} \\
		    &\qquad -\frac{{\mathds{G}_{ki}^{\rm E}}^\dagger\Big(\boldsymbol{D}\blockt\boldsymbol{A}\Big)_{jk}^\dagger}{2\ri}\Bigg\} \\
		    &\qquad +\frac{\mathds{G}_{ij}^{\rm E}-{\mathds{G}_{ji}^{\rm E}}^\dagger}{2\ri}\Bigg]. 
\end{split}
\label{gl2}
\end{equation}

Now setting Eq.~(\ref{gl1}) equal to Eq.~(\ref{gl2}) in thermal equilibrium and
 using block matrizes with $(\mathds{F})_{ij}=\mathds{F}_{ij}$, so of course $(\mathds{F}^\dagger)_{ij}=(\mathds{F}_{ji})^\dagger$ for $\mathds{F}\in\{\blockt, \boldsymbol{A}, \boldsymbol{D}, \mathds{G}^{\rm E}\}$ and remind that $\mathds{F}_{ij}^\dagger\coloneqq(\mathds{F}_{ij})^\dagger$  we obtain Eq.~(\ref{eqrelation}) 
\begin{eqnarray}
	\chi\blockt(\blockt)^\dagger&=&-k_0^2|\alpha|^2\blockt\IM\mathds{G}^{\rm E}(\blockt)^\dagger \notag \\
  & &+\frac{\blockt\alpha-(\blockt)^\dagger \alpha^*}{2\ri}.
\end{eqnarray}
Note that when taking the trace of the diagonal components of the block matrix then this relation reproduces the relation in Eq.~(29) in Ref.~\cite{nteilchen} which was derived by the requirement that in a $N$-dipole setup the heat emitted or received by any particle must vanish in equilibrium. Hence, our general relation reproduces a more specific result. Furthermore, it should also be noted that when taking the case of only one particle in vacuum ($\blockt = \mathds{1}$ and $\Im(\mathds{G}_{11}^{\rm E}) = k_0/6 \pi \mathds{1}$) our relation simply yields the {form of the suceptibility $\chi$ needed to have a consistent theory which is given by}  
\begin{equation}
  \chi = \Im(\alpha) - \frac{k_0^3}{6 \pi} |\alpha|^2
\label{Eq:Ci}
\end{equation}  
and was derived in exactly this way in Ref.~\cite{nteilchen}, but for {a single} particle only.

\section{{Breakdown bulk-edge correspondence}} \label{bbec}
{
In this appendix, we shortly discuss the possibility to observe the breakdown of the bulk-edge correspondence~\cite{PocockEtAl2019} in the
energy density above the 1D SSH chain. To this end, in analogy to the study in Ref.~\cite{PocockEtAl2019} we rewrite the eigenvalue Eq.~(\ref{Eq:SSH1D})
as
\begin{equation}
	\tilde{\mathds{M}}^{\nu} \begin{pmatrix} p^{\nu}_A \\ p^\nu_B \end{pmatrix} = \frac{d^3}{\alpha}\begin{pmatrix} p^{\nu}_A \\ p^\nu_B \end{pmatrix} = E \begin{pmatrix} p^{\nu}_A \\ p^\nu_B \end{pmatrix} 
\label{Eq:SSH1Db}
\end{equation}
with $	\tilde{\mathds{M}}^{\nu} = \mathds{M}^{\nu} d^3$. The eigenvalues $E$ of this equation allow to identify the lattice constant $d$ for which 
the edge modes enter the bulk bands. We furthermore use the same approximations as in Ref.~\cite{PocockEtAl2019} by focusing on $\omega = \omega_{\rm LP}$. This happens for transversal but not longitudinal modes indicating the breakdown for the transversal modes only~\cite{PocockEtAl2019}.}

\begin{figure}[h!]
	\centering
	\includegraphics[width=0.4\textwidth]{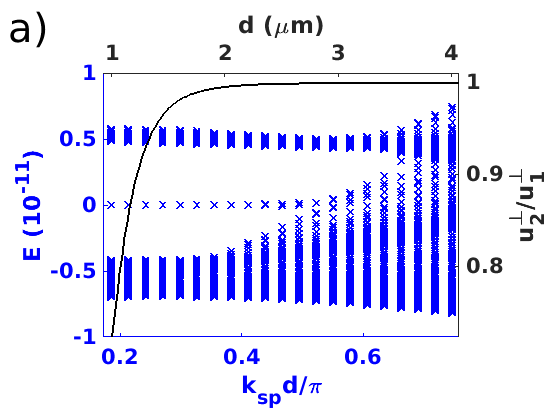}\\
        \includegraphics[width=0.4\textwidth]{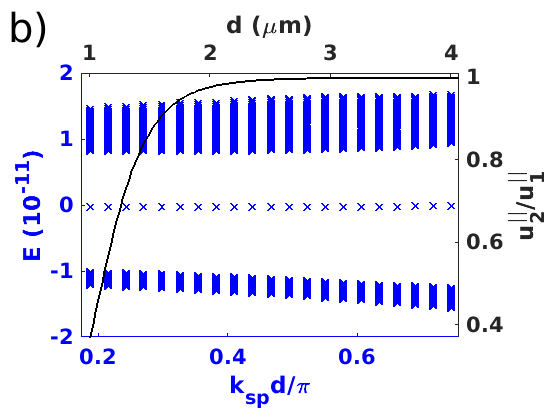}
		\caption{{Real part of the eigenvalues $E$ for a) transversal and b) longitudinal modes as function of lattice constant for $N = 300$ NPs and $\beta = 1.3$. Furthermore the ratio $u_2^\nu/u_1^\nu$ with $\nu = \perp, \parallel$ for the energy density 300nm above the second NP $u_2$ and the first NP $u_1$.}}
	\label{retardierung}
\end{figure}

{In Fig.~\ref{retardierung} we show the band structure in terms of the eigenvalues $E$ as a function of the lattice constant $d$. It can be seen that for the transversal modes the edge modes around $E = 0$ enter the bulk bands at $k_{\rm sp} d / \pi \approx 0.55$ ($k_{\rm sp} = \omega_{\rm LP}/c$) which is in exact agreement with the value found in Fig.~3c) in Ref.~\cite{PocockEtAl2019}. It corresponds to a lattice constant of $d \approx 3\,\mu{\rm m}$. In addition, in Fig.~\ref{retardierung} we have plotted the energy density ratio $u_2^\nu / u_1^\nu$ of the energy density $u_2^\nu$ ($u_1^\nu$) at $z = 300\,{\rm nm}$ above the second (first) NP. One can observe that for both the transversal and longitudinal modes this ratio converges to one at smaller values of $k_{\rm sp} d / \pi \approx 0.5$ corresponding to $d = 2.5\,\mu{\rm m}$. Hence, the energy density does not give a clear signal anymore for the edge modes, which is due to the fact that the coupling between the particles becomes very weak for such distances and the bands are getting very narrow around $\omega_{\rm LP}$. This means that in our configuration the breakdown of the bulk-edge correspondence cannot be seen with near-field thermal microscopes even if it would be possible to separate the contributions of the transversal and longitudinal modes to the energy density.}

\end{document}